\documentclass[aps,prd,amsmath,amssymb,10pt,letterpaper,balancelastpage,showpacs,superscriptaddress,nofootinbib,floatfix,notitlepage,longbibliography,twocolumn
]{revtex4-2}
\usepackage[utf8]{inputenc}
\usepackage[normalem]{ulem}
\usepackage{comment}
\usepackage{graphics}
\usepackage{bm}
\usepackage{graphicx}
\usepackage{color}
\usepackage{amsmath}
\usepackage{amssymb}
\usepackage{mathrsfs}
\usepackage{graphicx}
\usepackage{caption}
\usepackage{subcaption}
\usepackage{accents}
\usepackage{adjustbox}
\usepackage{multirow}
\usepackage{diagbox}
\usepackage{stackengine}
\usepackage{float}
\usepackage{ulem}

\usepackage[dvipsnames]{xcolor}
\definecolor{red}{rgb}{0.9, 0,0}
\definecolor{cerulean}{rgb}{0., 0.42,0.9}
\definecolor{navy}{rgb}{0.05, 0.05,0.8}

\usepackage[colorlinks]{hyperref}
\hypersetup{
  colorlinks = true,
  citecolor = red,
	linkcolor = navy
}
\usepackage{slashed}

\newcommand{\eq}[1]{Eq.~(#1)}
\newcommand{\Eq}[1]{Eq.~(#1)}

\newcommand{\Lagrangian}{\mathcal{L}}
\newcommand{\rhoDM}{\rho_{\text{DM}}}

\newcommand{\vbold}{\mathbf{v}}

\newcommand{\vmax}{v_{\text{max}}}

\newcommand{\tpres}{t_{\rm pres}}
\newcommand{\tperi}{t_{\rm peri}}
\newcommand{\GxA}{\Gamma_{\chi A}}
\newcommand{\vmin}{v_{\text{min}}}
\newcommand{\sxn}{\sigma_{\chi n}}

\newcommand{\muA}{\mu_A}
\newcommand{\nx}{n_{\chi}}
\newcommand{\Ht}{\tilde{H}}
\newcommand{\mx}{m_{\chi}}
\newcommand{\mrock}{M}
\newcommand{\Gxrock}{\Gamma_{\chi,\rm mineral}}
\newcommand{\RxA}{R_{\chi A}}
\newcommand{\kms}{\text{km}/\text{s}}
\newcommand{\vesc}{v_{\text{esc}}}
\newcommand{\Trock}{T}

\newcommand{\SNR}{\text{SNR}}

\newcommand{\mH}{m_{\Ht}}
\newcommand{\gpg}{\text{g}/\text{g}}

\begin{document}
\title{Heavy-element paleodetectors for Higgsino dark matter}

\author{Peter W.~Graham}
\affiliation{Leinweber Institute for Theoretical Physics, Department of Physics, Stanford University, Stanford, CA 94305, USA}
\affiliation{Kavli Institute for Particle Astrophysics and Cosmology, Department of Physics, Stanford University, Stanford, CA 94305, USA}

\author{Harikrishnan Ramani}
\affiliation{Department of Physics and Astronomy, University of Delaware and the Bartol Research Institute, Newark, DE 19716, USA}

\author{Samuel S.~Y.~Wong}
\email{samswong@uw.edu}
\affiliation{Department of Physics, University of Washington, Seattle, WA 98195, USA}

\begin{abstract}
Paleodetectors have been proposed as a new approach to direct detection of weakly interacting massive particles (WIMPs), through the search for damage tracks in ancient minerals induced by WIMP-nucleon scattering. However, for inelastic dark matter such as the Higgsino, existing paleodetector targets lack sufficiently heavy nuclei to overcome the kinematic threshold for scattering. We propose heavy-element paleodetectors as a new probe of inelastic dark matter, using ancient, radiopure minerals containing heavy elements such as lead. We identify brine precipitates from deep geothermal aquifers as a possible geological source of such minerals. Additionally, paleodetectors are uniquely sensitive to the history of the dark matter high-velocity tail, including a possible fast population induced by the Large Magellanic Cloud’s close approach 50 Myr ago. Such a scenario would favor younger minerals than usually assumed in the paleodetector literature. This method can probe Higgsino mass splittings up to $\delta \simeq 920~\text{keV}$. Due to the large Higgsino-nucleon cross section, we find that even suboptimal mineral samples with ordinary radioactivity from depths of only 2~km can probe new parameter space, thus partially relaxing the stringent requirements on radiopurity and depth that constitute two significant challenges for the paleodetector program. 
\end{abstract}

\maketitle

\tableofcontents

\section{Introduction}

Identifying the particle nature of dark matter (DM) is one of the primary goals of contemporary particle physics. Among many models, the weakly interacting massive particle (WIMP) paradigm stands out as particularly well-motivated: a particle with mass and annihilation cross section roughly the size of the electroweak scale would be thermally produced with a relic density consistent with the one observed for dark matter today. This ``WIMP miracle" has motivated many direct, indirect, and collider searches probing the WIMP window. In particular, underground xenon detector experiments have placed stringent limits on the WIMP-nucleon elastic cross section~\cite{LZ:2022lsv,DarkSide-50:2022qzh,XENON:2023cxc,PandaX:2024qfu}.

In recent years, paleodetectors have been proposed as an alternative method for WIMP searches~\cite{paleodetectors_2018,paleodetectors_details_2018,paleodetectors_spectral_2018,paleodetectors_new_2021,paleo_white_paper_2023,paleodetectors_refine_2025,Calabrese-Day:2026soq}, reviving an old idea from Ref.~\cite{Mica_Snowden_1995}. If a WIMP scatters with a nucleus in a mineral, the recoiling nucleus would produce a damage track in the material. A large number of these DM-induced tracks would accumulate in certain ancient minerals, up to $\mathcal{O}(\text{Gyr})$ in age, and would be excellent search targets given sufficiently low background. Even a small volume of such minerals could have exposure that exceeds that of a conventional xenon experiment. Paleodetectors have been projected to probe elastic cross sections smaller than current limits $\sim 10^{-46}~\text{cm}^2$~\cite{paleodetectors_new_2021}.

A particularly well-motivated class of WIMPs is electroweak WIMPs, in which dark matter interacts through \emph{the} standard model (SM) electroweak interaction, rather than through new interactions of the same approximate
strength~\cite{LastWIMP}. An electroweak WIMP with nonzero hypercharge generically scatters with nucleons through tree-level $Z$ exchange, with a cross section $\simeq 10^{-39}~\text{cm}^2$.
While such a large elastic cross section has long been ruled out by xenon experiments,
one remaining possibility is the inelastic dark matter scenario~\cite{Tucker-Smith:2001myb}, in which a small mass splitting $\delta$ exists between two nearly degenerate states that only couple to the $Z$ boson off-diagonally. Then, if $\delta$ is larger than the available incoming kinetic energy, the scattering would be kinematically forbidden. Furthermore, this scenario is realized in supersymmetry (SUSY) if the DM candidate is the Higgsino, thus making it one of the last classic supersymmetric WIMPs standing~\cite{LastWIMP}.

Inelastic DM with mass splittings $\delta \gtrsim 350$~keV would be invisible to both xenon experiments and existing paleodetector proposals. The reason is that the kinematic threshold set by the mass splitting $\delta$ can only be overcome by a sufficiently heavy target nucleus. The current kinematic reach is set by the xenon mass, whereas the heaviest element in existing paleodetectors is iron~\cite{Theodosopoulos:2026ehn}, which is lighter than xenon.

To overcome this obstacle, we propose a variant of paleodetectors that we term ``heavy-element paleodetectors." Our approach targets inelastic DM by using appropriate, radiopure minerals that contain heavy elements such as lead (Pb). In particular, we find that such a detector can cover large parts of previously unexplored inelastic DM parameter space.  And we identify a geological formation mechanism that, unlike those previously considered in the paleodetector literature, can produce radiopure minerals containing heavy elements: mineral precipitation in deep underground aquifers of geological age containing geothermal brines with low concentrations of uranium and high concentrations of Pb.

Interestingly, due to the interaction of the Large Magellanic Cloud (LMC) with the Milky Way (MW), the DM velocity tail may have been faster in the past than it is today~\cite{LMC}.  Paleodetectors are uniquely sensitive to this history. The history of the velocity tail does not necessarily favor the oldest minerals, contrary to usual assumptions, but instead favors those that are just old enough to reach the period with the highest velocity tail. Since there is still significant uncertainty surrounding this astrophysical scenario, we will present our results both for this LMC model for the velocity distribution as well as for the Standard Halo Model (SHM).

The large Z-mediated cross section opens up the exciting possibility of discovering dark matter with suboptimal but much more accessible mineral samples: radio-impure samples from existing underground laboratories at depths of $2~\text{km}$. In contrast, radiopure samples from $5~\text{km}$ depth are required both for probing the largest mass splittings in inelastic DM and for conventional paleodetector searches. Thus, inelastic DM could serve as an intermediate physics goal for the paleodetector paradigm.

While our proposal is sensitive to generic inelastic DM, for concreteness, we will take the Higgsino with a relic mass of 1.1~TeV as a well-motivated benchmark particle. We now briefly summarize its current experimental status~\cite{EnhancingHiggsinoDM_2024}. The relic-mass Higgsino lies well beyond existing collider limits~\cite{aaboud2018search} and the reach of even future 100-TeV colliders~\cite{Low:2014cba}. There are interesting prospects for its indirect detection~\cite{Rinchiuso:2020skh,Rodd:2024qsi,Abe:2025lci} at the CTA~\cite{CTAConsortium:2010umy} in the near future, but they are subject to significant uncertainty in the DM density at the galactic center. Electron electric dipole moment (EDM) experiments such as ACME II~\cite{acme2018improved,Co:2021ion} also provide complementary probes by setting limits on gaugino masses (see Section~\ref{sec: Higgsino model}), $M_{1,2} \gtrsim 10$~TeV, corresponding to upper limits on the Higgsino mass splittings, $\delta \lesssim 200$~MeV; projections from Advanced ACME can improve this to $\delta \lesssim 30$~MeV.

In direct detection, PandaX-4T~\cite{PandaX4T2021} sets the current lower limit on Higgsino mass splitting, $\delta_{\text{SHM}} \gtrsim 260~\text{keV}$~\cite{EnhancingHiggsinoDM_2024} (assuming the SHM), beyond which is the ``inelastic frontier"~\cite{inelastic_frontier_Bramante_2016} that would require a target nucleus mass heavier than xenon or DM velocities faster than the MW escape velocity. Ref.~\cite{Luminous} proposes luminous detection~\cite{Feldstein:2010su} of Higgsinos, in which the Higgsinos could upscatter against the trace abundance of heavy elements in the Earth, and subsequently decay back to the lighter states via photons, which could be detected by large-volume underground neutrino detectors such as JUNO~\cite{JUNO_Conceptual}.

In Ref.~\cite{EnhancingHiggsinoDM_2024}, we showed that taking into account the DM velocity tail due to the LMC improves existing limits ($\delta_{\text{LMC}} \gtrsim 340~\text{keV}$) and luminous detection prospects. Furthermore, we showed that one could take advantage of the fast DM by placing a large volume of lead or depleted uranium near the neutrino detectors to enhance upscattering. In this work, we consider heavy-element paleodetectors as an alternative route toward pushing the inelastic frontier, which involves distinct technological challenges with unrivaled sensitivity to dark matter velocities scanned over the entire history of the Earth.

The rest of the paper is organized as follows. In Section~\ref{sec:inelastic_higgsino}, we review the phenomenology of inelastic DM as well as the Higgsino model. In Section~\ref{sec:conventional_paleodectors}, we review basic aspects of the existing paleodetector proposal; readers familiar with paleodetectors can skip to Section~\ref{sec:heavy_element_paleodetectors}, where we describe our proposed variant and qualitative results. In Section~\ref{sec: DM_signal}, we show the DM signal calculation, and the resulting projected limits are shown in Section~\ref{sec: projection}. We conclude in Section~\ref{sec: conclusion}.

\section{The inelastic frontier}
\label{sec:inelastic_higgsino}

We begin by reviewing inelastic dark matter as a loophole of the direct detection paradigm (Sec~\ref{sec:inelastic_kinematics}), and how heavy elements and a possible DM high-velocity tail (Sec~\ref{sec:LMC}) could push this inelastic frontier. We provide a brief review of the Higgsino as a well-motivated realization (Sec~\ref{sec: Higgsino model}).

\subsection{Inelastic kinematics}
\label{sec:inelastic_kinematics}

\begin{figure}[t]
    \centering
    \includegraphics[width=0.5\textwidth]{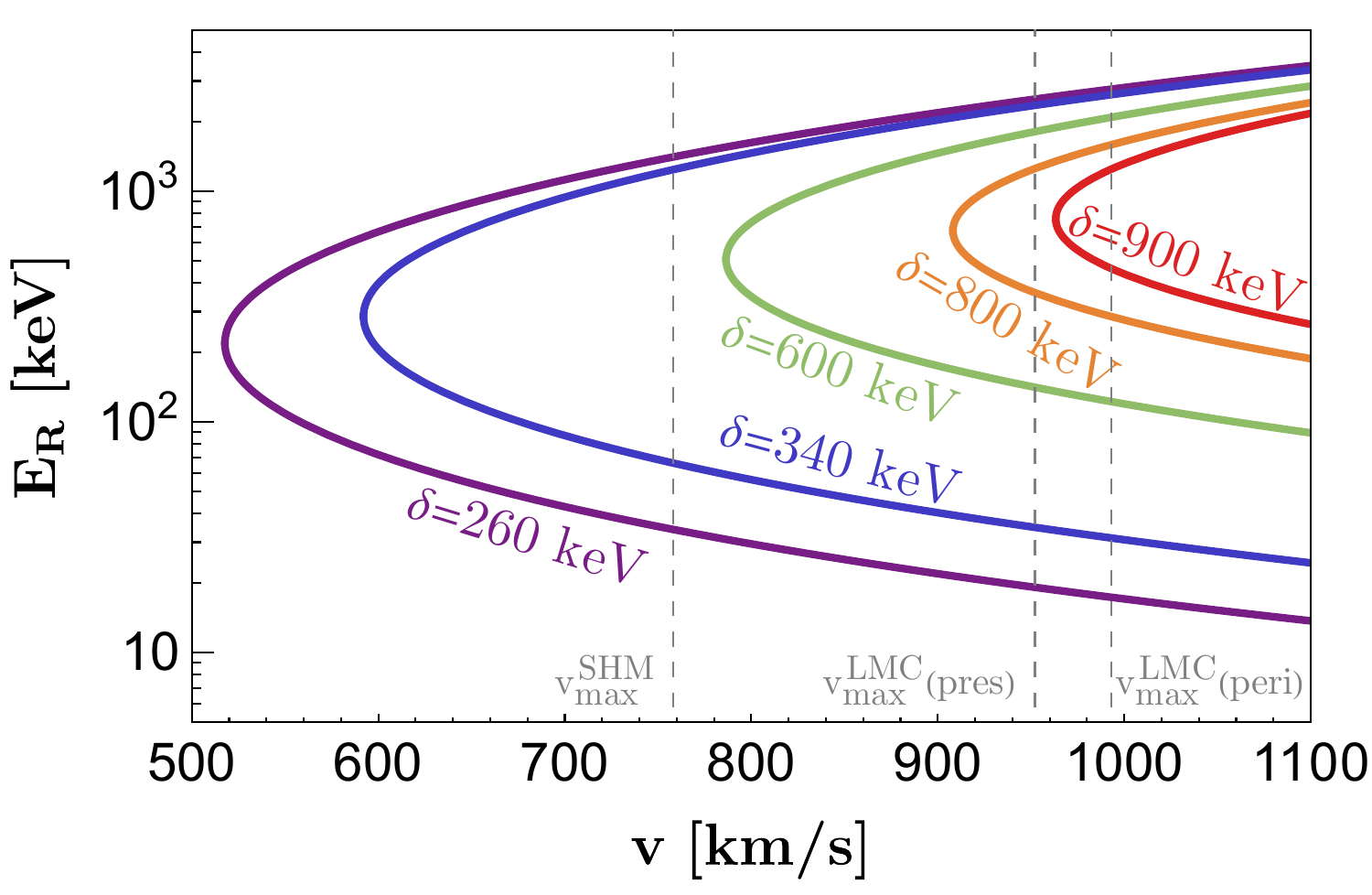}
    \caption{Contours of recoil energy $E_R$ of Pb nuclei as a function of DM velocity $v$ for various mass splittings $\delta$. Maximum DM velocities of various models are shown in dashed vertical lines.}
    \label{fig: ER contour}
\end{figure}

Consider an incoming DM particle $\chi$ with lab-frame velocity $v$ and mass splitting $\delta$ scattering with a nucleus of mass number $A$ and mass $m_A=A m_n$. The nuclear recoil energy $E_R$ is~\cite{inelastic_frontier_Bramante_2016}
\begin{align} \label{eq: ER full euqation}
    E_R &= \frac{\mu_A}{m_A} \left[(\mu_A v^2 \cos^2 \theta - \delta ) \right. \nonumber \\
    &\pm (\mu_A v^2 \cos^2 \theta)^{1/2} (\mu_A v^2 \cos^2 \theta - 2 \delta )^{1/2} \left. \right] ~,
\end{align}
where $\mu_A=\mx m_A/(\mx + m_A)$ is the DM-nucleus reduced mass, and $\theta$ is the lab-frame scattering angle. For a given $E_R$, the minimum $v$ occurs at extremal scattering angles $\cos^2 \theta =1$,
\begin{align} \label{eq: vmin}
    \vmin(E_R) = \frac{1}{\sqrt{2m_A E_R}}\left( \frac{m_A}{\mu_A}E_R + \delta \right) ~.
\end{align}

For each $\delta$, $\vmin(E_R)$ traces a contour in the $E_R$-$v$ plane, shown in Fig.~\ref{fig: ER contour} for Pb nuclei and $\mx=1.1$~TeV. The region consistent with upscattering is bounded on the left by the contour and on the right by the maximum DM velocity $\vmax$ (vertical dashed lines; see Section~\ref{sec:LMC}).
The apex of each contour occurs at $v_{\rm min}^{\rm apex}=\sqrt{2\delta/\mu_A}$,
so that the allowed kinematic space shrinks as $\delta$ increases, and the maximum $\delta$ we can probe is
\begin{align} \label{eq:delta_max}
    \delta_{\rm{max},A} = \frac{1}{2} \mu_A \vmax^2 ~.
\end{align}

\Eq{\ref{eq:delta_max}} demonstrates how inelastic DM is a loophole in direct detection. If $\delta > \delta_{\rm{max},A}$, the DM-nucleon scattering is kinematically forbidden by conservation of energy. Hence, inelastic DM with large cross sections that would have otherwise been detected can evade conventional direct detection. This is the reason electroweak WIMPs interacting via tree-level Z exchange can still be viable.

\Eq{\ref{eq:delta_max}} also provides two clear paths toward pushing the inelastic frontier. Since $\mu_A \approx m_A$ for $m_{\chi} \gg m_A$, sensitivity to $\delta$ is directly proportional to the mass of the target nucleus. Current limits on $\delta$ by xenon experiments are set by the xenon mass ($A_{\rm Xe}=131$). Heavy elements such as lead ($A_{\rm Pb}=207$) are thus powerful tools for probing DM inelasticity. Finding paleodetectors containing heavy elements is the main point of this paper. The other way to increase $\delta_{\rm{max},A}$ is to target the high-velocity tail of the DM distribution, which we turn to next.

\subsection{Ancient, fast dark matter from the LMC}
\label{sec:LMC}

\begin{figure}[t]
    \centering
    \includegraphics[width=0.5\textwidth]{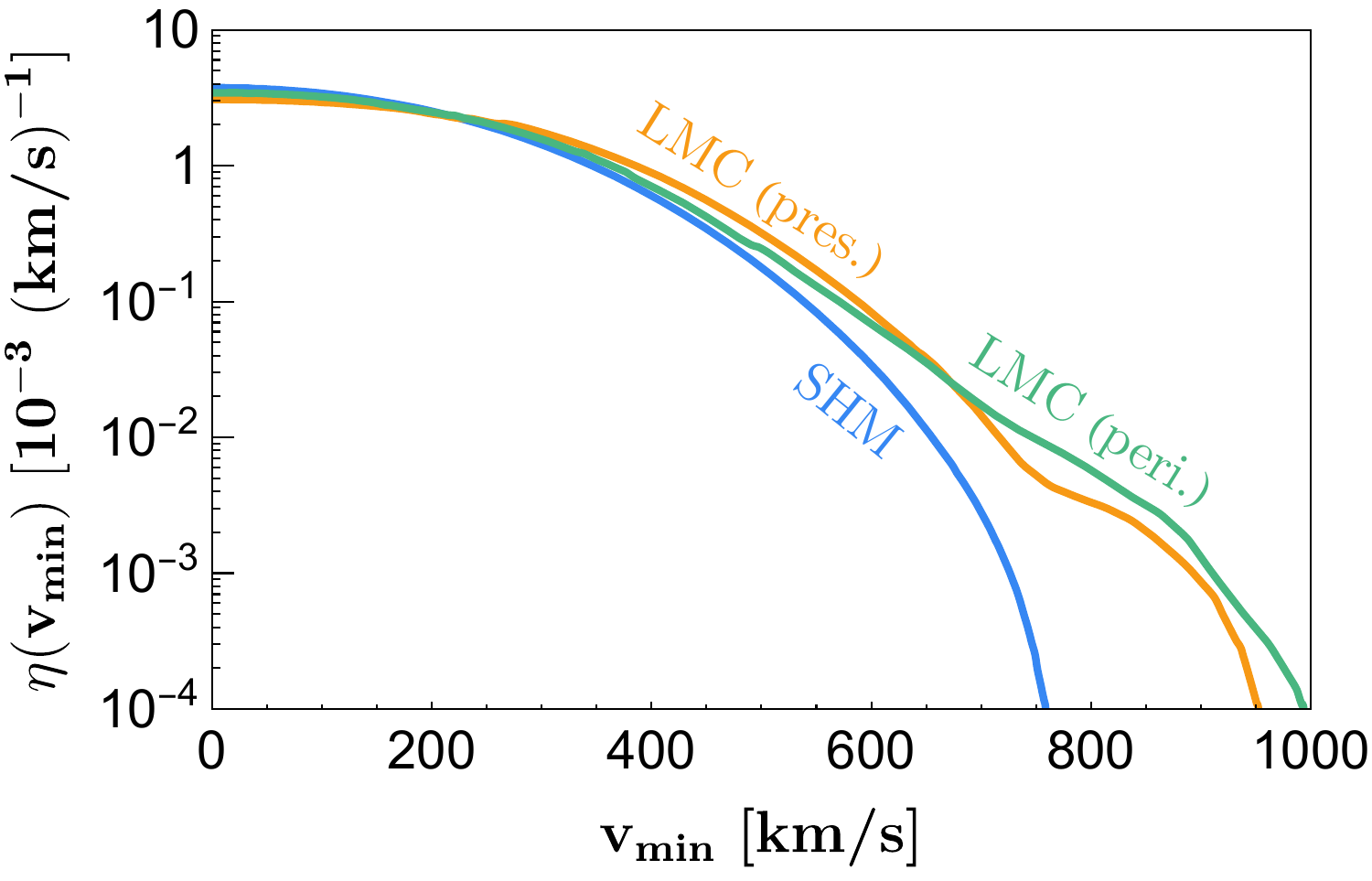}
    \caption{Halo integrals of the SHM and LMC model (at present and at pericenter). Figure adapted from Ref.~\cite{LMC}.}
    \label{fig: halo}
\end{figure}

Conventionally, most analyses assume the Standard Halo Model (SHM)~\cite{Original_SHM}, which is a truncated Maxwell-Boltzmann distribution in the galactic frame,
\begin{align}
    f^{\text{gal}}_{\text{SHM}}(\vbold^{\text{gal}}) ~ \propto ~ e^{
    -\left(v^{\text{gal}}/v_0^{\text{gal}}\right)^2
    }
    \Theta(\vesc^{\text{gal}} - v^{\text{gal}}) ~,
\end{align}
where $v_0^{\text{gal}}=220 ~\kms$ is the local standard of rest,  $\vesc^{\text{gal}}=540~\kms$ is the MW escape velocity~\cite{LMC}, and $\Theta$ is the Heaviside step function.

However, recent work~\cite{LMCHighSpeed, LMC, LMC_Similar} in galactic dynamics shows that the LMC's close pericenter approach to the MW $(\tpres - \tperi) \sim 50$~Myr ago~\cite{LMC_pericenter_time} in the opposite direction to solar motion may have resulted in an appreciable population of fast, unbound DM in the Earth's frame. This effect was more pronounced at the time of the pericenter approach, but due to the relative recency of this encounter, present-day DM has not fully equilibrated and still retains high velocities. Its implications for direct detection of inelastic dark matter were first explored by us in Ref.~\cite{EnhancingHiggsinoDM_2024}. In this work, we show that paleodetectors are uniquely sensitive to the possibility of even faster DM in the past.

Ref.~\cite{LMC} modeled the DM velocity distribution taking into account the interaction and formation history of the MW and LMC using the Auriga cosmological magneto-hydrodynamical simulations~\cite{Auriga}. For both the SHM and this LMC model\footnote{This model is called halo 13 in Ref.~\cite{LMC}, also known as Auriga halo 25.}, we take from Ref.~\cite{LMC} the mean inverse velocity, or the halo integral (see Fig.~\ref{fig: halo}), defined as
\begin{align} \label{eq:halo integral}
    \eta(\vmin,t) \equiv \int_{v>\vmin} d^3v ~ \frac{1}{v}f^{\text{lab}}(\vbold,t) ~,
\end{align}
where $f^{\text{lab}}(\vbold,t)$ is the lab-frame DM velocity distribution, and $\vmin$ is the minimum velocity required for a particular scattering. The time dependence on the scale of a year due to the Earth's orbit around the Sun has been averaged out, but the longer timescale due to the LMC is captured by the two snapshots, at present and at pericenter. Table~\ref{tab: halo model properties} demonstrates that the higher characteristic velocities of the LMC, together with the larger mass of nuclei like Pb, extend the reach in $\delta$ beyond the reach of SHM and iron (Fe), the heaviest element in conventional paleodetectors~\cite{Theodosopoulos:2026ehn}.

\begin{table}[t]
    \centering
    \renewcommand{\arraystretch}{1.2} 
    \begin{tabular}{c| c  c  c}
        \hline
        halo model & $\vmax$~[\text{km/s}] & $\delta_{\rm{max}}^{\rm Pb}$~[\text{keV}] & $\delta_{\rm{max}}^{\rm Fe}$~[\text{keV}] \\  
        \hline
        SHM & 760 & 560 & 170 \\ 
        LMC (present) & 950 & 880  & 270 \\ 
        LMC (pericenter) & 990 & 960  & 290 \\ 
        \hline
    \end{tabular}
    \caption{The maximum DM lab-frame velocity and maximum splitting $\delta$ in different halo models, for lead (Pb) and iron (Fe) nuclei.}
    \label{tab: halo model properties}
\end{table}

We will present our results both for this LMC model for the velocity distribution as well as for the SHM, since there is still significant uncertainty surrounding this possibility of a high-velocity tail.

In this work, we restrict to the LMC merger, conservatively assuming the SHM remained valid until $\sim 50 $~Myr ago.
Even though recent studies \cite{LMCHighSpeed, LMC, LMC_Similar} argue that the present-day high-velocity tail predominantly originates from the LMC, this still leaves open the question of whether earlier mergers enhanced the tail at earlier times, to which paleodetectors are uniquely sensitive. Up to this point, much of the interest in the literature has focused on the present-day DM velocity distribution~\cite{SMH_Deviation_2,Mao:2012hf,SMH_Deviation_5,Herzog-Arbeitman:2017fte,Necib:2018igl,Zhang:2026qnl}, whereas our work points to a new direction by motivating the study of the history of the DM velocity tail. We leave this for future work and encourage the astrophysics community to study it.

\subsection{Higgsino dark matter}
\label{sec: Higgsino model}

While this work applies to generic inelastic DM, we take the Higgsino, one of the last classic supersymmetric WIMPs standing, as a well-motivated benchmark particle. We briefly review this model below, which has been studied extensively elsewhere~\cite{Bhattiprolu:2025zwt,Co:2021ion,Kowalska:2018toh,LastWIMP}.

In the minimal supersymmetric standard model (MSSM), the up- and down-type Higgsino doublets together form a Dirac fermion with mass $\mu$ before electroweak symmetry breaking (EWSB). However, after EWSB, dimension-five operators involving the SM Higgs split the neutral Dirac Higgsino into two Majorana fermions~\cite{Nagata:2014wma}. Upon diagonalizing the mass matrix, the two neutral components $\Ht_1$ and $\Ht_2$ have masses $M_{\Ht_{1,2}} \approx \mu \mp \delta/2$, where $\delta$ is the Majorana mass given by
\begin{align}
\delta \approx m_Z^2\left(\frac{\sin^2\theta_W}{M_1}+\frac{\cos^2\theta_W}{M_2}\right) ~,
\label{eq:splitting}
\end{align}
with $\theta_W$ being the Weinberg angle, and $m_Z$, $M_1$, and $M_2$ being the Z boson, bino, and wino masses, respectively.

In the heavy-gaugino limit ($M_{1,2} \gg m_Z, \mu$), the Higgsinos are narrowly-split pseudo-Dirac fermions, with a small mass splitting $\delta \ll \mu$. The possibility that many of the superpartners are much heavier than the weak scale has attracted attention in light of the absence of new physics at the LHC~\cite{Nagata:2014wma,split_susy,split_susy_conditions,split_susy_little_hierarchy_2003,Gonski:2025wzh}.

If the Higgsino is the lightest supersymmetric particle (LSP)~\cite{LSP_1984} with mass $\mu \approx 1.1$~TeV, it would be produced via thermal freeze-out with the observed DM density $\Omega h^2 = 0.12$~\cite{Kowalska:2018toh}. Present-day DM would then be entirely made of the lighter state\footnote{We are dropping the subscript of $\Ht_1$ wherever there is no ambiguity.} $\Ht$ (of mass $\mH \equiv M_{\Ht_1}$), since the lifetime of the heavier state $\Ht_2$ (of mass $\mH + \delta$) is much shorter than the age of the universe.

The Higgsino scatters with nucleons inelastically via the off-diagonal coupling between $\Ht_1$ and $\Ht_2$ to the Z boson~\cite{Luminous,Tucker-Smith:2001myb,Bottaro:2022one}.
The Z-mediated Higgsino-nucleon cross section is~\cite{Kowalska:2018toh} 
\begin{align}
    \sigma_{n\Ht} \approx \frac{G_F^2m_n^2}{8\pi} \approx 10^{-39}~\text{cm}^2~,
\end{align}
many orders of magnitude larger than elastic cross sections already ruled out by direct detection. However, due to its inelastic nature, the interaction is kinematically forbidden if the mass splitting $\delta$ is sufficiently large.

Elastic scattering between Higgsinos and nucleons could occur via tree-level Higgs exchange or one-loop processes. Since the coupling between the Higgsino and the Higgs occurs via mixing with the gauginos, it is suppressed by the ratio $m_Z/M_{1,2}$ in the heavy-gaugino limit~\cite{LastWIMP}. On the other hand, elastic scattering via one-loop processes is highly suppressed due to an accidental cancellation between different diagrams at the observed Higgs boson mass of 125~GeV~\cite{Hisano:2011cs}, such that the cross section lies beneath the neutrino floor\footnote{Elastic scattering is also suppressed for other electroweak WIMPs, but the cross sections are somewhat larger than that of the Higgsino, though detecting them with conventional xenon detectors would still require huge improvements in detector volumes~\cite{Bottaro:2022one}.}~\cite{Hisano:2012wm}.

\section{Conventional paleodetectors}
\label{sec:conventional_paleodectors}
Xenon direct-detection experiments such as PandaX~\cite{PandaX4T2021}, XENON~\cite{XENON1T}, and LZ~\cite{LZ:2019sgr} aim to maximize DM exposure by using a ton-scale target mass for an integration time of $\mathcal{O}(1-10)$~years. Paleodetectors have been proposed as an alternative strategy~\cite{paleodetectors_2018,paleodetectors_details_2018,paleodetectors_spectral_2018,paleodetectors_new_2021,paleodetectors_refine_2025,Theodosopoulos:2026ehn}, reviving an old idea from Ref.~\cite{Mica_Snowden_1995}, in which the exposure is maximized via a long integration time on geological scales.

If a WIMP particle scatters with a nucleus in a mineral, the recoiling nucleus could produce a damage track in the material. A large number of these DM-induced tracks would accumulate in certain ancient minerals, up to $\sim $Gyr in age, and would be excellent search targets given sufficiently low background. Even a small mass of such a mineral could have exposure exceeding that of xenon experiments; for instance, a 100~g Gyr-old sample has an exposure of $10^5$~ton-years.

We briefly review the existing paleodetector proposal based on Refs.~\cite{paleodetectors_2018,paleodetectors_details_2018} in this section; readers familiar with paleodetectors can skip to Section~\ref{sec:heavy_element_paleodetectors}.

\subsection{Read-out methods}
\label{sec: readout}
To record tracks, materials must be insulators or poor semiconductors with electrical resistivity\footnote{Note that the exact mechanism for track formation is not fully understood, so this requirement should be considered a rule of thumb rather than a hard rule.}~\cite{paleodetectors_details_2018,GUO2012233}
\begin{align} \label{eq:resistivity_requirement}
    \rho_r \gtrsim 2000~\Omega~\text{cm} ~.
\end{align}
Such materials are called solid-state track detectors (SSTDs)~\cite{Fleischer:1964,Fleischer383,Fleischer:1965yv,GUO2012233}.
For the tracks to be preserved, the annealing time, which depends on the material’s melting point relative to the temperature of its environment, must be longer than the age of the sample~\cite{paleodetectors_2018,paleodetectors_details_2018}.

Three-dimensional reconstruction of damage tracks from mineral samples could potentially be achieved via one of the following read-out methods~\cite{paleodetectors_2018,paleodetectors_details_2018}: 
\begin{itemize}
    \item Helium Ion Beam Microscopy (HIBM)~\cite{Hill:2012} could achieve high spatial resolution of $\sigma_x \approx 1$~nm, but its relatively low efficiency only allows for processing a target volume\footnote{The paleodetector literature quotes benchmark target masses, which we convert to target volumes using the density of Epsomite. For comparing read-out capacities, the relevant variable to be held fixed should be the target volume, which translates to larger target masses for heavy-element minerals due to their larger densities.} of $6~\text{mm}^3$.
    \item Small Angle X-ray scattering (SAXs) tomography at synchrotron facilities~\cite{Schaff:2015} has a lower spatial resolution of $\sigma_x \approx 15$~nm~\cite{SAX_resolution_2014}, but its higher efficiency allows for processing a target volume of $60~\text{cm}^3$.
\end{itemize}
In general, WIMPs with masses less than $10$~GeV create shorter tracks and are better targeted by HIBM, while higher-mass WIMPs that induce longer tracks can take advantage of SAXs.

\subsection{Background mitigation}
Several sources of background need to be mitigated~\cite{paleodetectors_2018,paleodetectors_details_2018}. First, mineral samples need to be obtained from depths greater than $\sim 5$~km for the rock overburden to sufficiently shield cosmic-ray-induced neutrons.

Second, typical minerals in the Earth's crust have prohibitively large concentrations of radioactive elements, particularly $^{238}\text{U}$, on the order of $\sim 10^{-6}~\gpg$ by weight. In contrast, Refs.~\cite{paleodetectors_2018,paleodetectors_details_2018,Baum:2019fqm}
point out that seawater and the Earth's mantle are naturally much more radiopure, and propose to use minerals in marine evaporites (MEs), formed at the bottom of evaporating oceans, and ultra-basic rocks (UBRs), formed in the Earth's mantle. The benchmark $^{238}\text{U}$ concentrations by weight are assumed to be $C^{238} = 10^{-11}~\gpg$ for MEs and $C^{238} = 10^{-10}~\gpg$ for UBRs~\cite{paleodetectors_2018,paleodetectors_details_2018}. Due to the range of fast neutrons, the radiopurity requirement applies not only to the target mineral itself, but also to its surrounding environment over a volume of $\mathcal{O}(\text{m}^3)$~\cite{Baum:2019fqm}.

Third, even these radiopure samples will contain a substantial number of $\alpha$ decays, whose recoiling daughter nuclei leave background tracks (whereas the emitted $\alpha$ particles generally do not, and neither do the products of $\beta$ or $\gamma$ decays)~\cite{Theodosopoulos:2026ehn}. But since $\alpha$ decays are two-body decays, with mono-energetic nuclear recoil, each radioactive decay has a characteristic track length in a given material. Furthermore, while $^{238}\text{U}$ has a half-life of $T^{238}_{1/2}=4.5 \times 10^9$~years, all of the subsequent decays are much faster, with the second-longest lived, $^{234}\text{U}$, having a half-life of  $T^{234}_{1/2}=2.5 \times 10^5$ years. Since the relevant minerals considered are much older than $10^5$ years, almost all $^{238}\text{U}$ nuclei that have undergone the first decay ($^{238}\text{U} \to ^{234}\text{Th}+\alpha$) will have proceeded through the entire decay chain to the stable $^{206}\text{Pb}$. Hence, we assume we can reject these connected $\alpha$-tracks with characteristic lengths during the analysis~\cite{paleodetectors_details_2018}.

\subsection{Irreducible backgrounds}
\label{sec:irreducible_background}
Even after taking these background mitigation strategies into account, three irreducible backgrounds remain. First, due to the aforementioned $\sim 10^5$-year half-life of $^{234}\text{U}$, a small fraction of nuclei in the sample inevitably have undergone only a single $\alpha$-decay, namely, $^{238}\text{U} \to ^{234}\text{Th}+\alpha$. For $T^{234}_{1/2}\lesssim \Trock \lesssim  T^{238}_{1/2}$, where $\Trock$ is the age of the mineral, the number of single-$\alpha$ (1$\alpha$) events per unit target mass is time-independent and given by~\cite{paleodetectors_details_2018}
\begin{align} \label{eq:1alpha_number}
    n_{1\alpha} \approx 10^9~\text{kg}^{-1} \left( \frac{C^{238}}{10^{-11}~\gpg}\right) ~.
\end{align}
Since the $^{234}\text{Th}$ daughter nucleus carries the characteristic energy $E_R = 72$~keV, these 1$\alpha$ backgrounds form a sharp peak in the track-length spectrum. It is possible that the $\sim 10~\mu\text{m}$ $\alpha$-tracks are observable in some materials, in which case the 1$\alpha$ backgrounds could be rejected entirely~\cite{paleodetectors_details_2018}. We conservatively assume $\alpha$-tracks are not observable in our target materials in this work.

A second irreducible background is fast neutrons from both spontaneous fission (SF) of $^{238}\text{U}$ and $(\alpha,n)$ reactions, in which $\alpha$ particles from $\alpha$-decays render a different nucleus unstable~\cite{paleodetectors_details_2018}. Note that the primary daughter nuclei from the SF events themselves can be easily rejected due to their rich decay chain. The number of fast neutron events per target mass is directly proportional to $C^{238}$ and $\Trock$.

Due to the similar masses of the neutron and the proton, fast neutrons lose energy especially efficiently when scattering off hydrogen nuclei, and stop after fewer collisions than they would off heavier nuclei. Hence, if the mineral contains hydrogen (H) in its chemical formula, the fast neutron background is significantly lower. Another rule of thumb is that minerals containing lithium (Li) or beryllium (Be) are poor paleodetectors due to having large $(\alpha,n)$ cross sections~\cite{paleodetectors_details_2018}.

Finally, neutrinos from the Sun, supernovae, and cosmic ray interactions in the atmosphere give rise to an irreducible nuclear recoil background. The number of neutrino events per target mass is directly proportional to $\Trock$. While the present-day astrophysical neutrino fluxes are accurately known, these values could have been different in the past. We follow Refs.~\cite{paleodetectors_details_2018,paleodetectors_2018} in conservatively taking a relative uncertainty of $\epsilon_{\nu}=100\%$ for the time-integrated number of neutrino-induced events; in contrast, the normalizations of fast neutrons and $1\alpha$ backgrounds could be much more accurately determined, and relative uncertainties of $\epsilon_{n}=\epsilon_{1\alpha}=1\%$ are assumed~\cite{paleodetectors_details_2018,paleodetectors_2018}. Note that this requires determining the uranium concentration and the number of $\alpha$-tracks down to 1\% accuracy; the uranium concentration can be measured reliably to $\mathcal{O}(10^{-15}~\gpg)$ using inductively coupled plasma mass spectroscopy~\cite{paleodetectors_details_2018}.

The neutrino-induced track spectrum is essentially identical across different materials, and the $C^{238}$-normalized neutron-induced track spectrum depends essentially only on whether the material contains H (assuming no Li or Be); see Fig.~2 of Ref.~\cite{paleodetectors_details_2018}. Hence, for this initial study, we take the neutrino and neutron spectra of Epsomite $\left[\text{MgSO}_4 \cdot 7(\text{H}_2\text{O})\right]$ from Ref.~\cite{paleodetectors_2018} and assume they apply to generic materials with H. 

\section{Heavy-element paleodetectors}
\label{sec:heavy_element_paleodetectors}
 
Conventional paleodetectors have been projected to probe elastic cross sections smaller than current limits~\cite{paleodetectors_new_2021} and inelastic cross sections competitive with existing limits from LZ~\cite{LZ:2023lvz} for small mass splittings, $\delta \lesssim 50$~keV~\cite{Theodosopoulos:2026ehn}. While these are valuable probes of generic WIMPs, they are not sensitive to inelastic dark matter with sufficiently large mass splittings, such as the viable Higgsino parameter space. As reviewed in Section~\ref{sec: Higgsino model}, the Higgsino's elastic cross section lies beneath the neutrino floor, beyond the capabilities of both xenon experiments and paleodetectors. The Higgsino also has a Z-mediated inelastic cross section $\sigma_{n\Ht} \approx 10^{-39}~\text{cm}^2$. Such a large cross section has been ruled out unless the mass splitting is sufficiently large, $\delta_{\rm SHM} \gtrsim 260$~keV ($\delta_{\text{LMC}} \gtrsim 340~\text{keV}$)~\cite{EnhancingHiggsinoDM_2024}. Similar conclusions apply more generally to electroweak WIMPs~\cite{Bottaro:2022one,luminous_complex_WIMP}.

Conventional paleodetectors lose sensitivity for mass splittings above $\delta \gtrsim 100$~keV simply because typical minerals in MEs and UBRs do not contain heavy elements~\cite{Theodosopoulos:2026ehn}\footnote{However, we do not wish to rule out such possibilities because in very rare circumstances, Galena (PbS) could form in MEs~\cite{Stewart1963MarineEvaporites}.}. For MEs, the reason is that the major elements dissolved in seawater are all light elements, with calcium ($A_{\rm Ca}=40$) being the heaviest among them~\cite{Stewart1963MarineEvaporites}\footnote{Notably, barium ($A_{\rm Ba}=138$) is a minor element that dissolves in seawater and appears in Barite ($\text{BaSO}_4$) as an ME~\cite{Stewart1963MarineEvaporites}. It is substantially heavier than other elements in MEs, though still not much heavier than xenon ($A_{\rm Xe}=131$).}. Similarly, Fe ($A_{\rm Fe}=56$) is the heaviest common element in the Earth's mantle and hence also in UBRs~\cite{wyllie1970ultramafic}; it is part of Olivine $\left[\text{Mg}_{1.6}\text{Fe}^{2+}_{0.4}(\text{Si}\text{O}_4)\right]$ in conventional paleodetectors~\cite{Theodosopoulos:2026ehn}, but is not sensitive to viable Higgsino parameter space; see Table~\ref{tab: halo model properties}. The typical lead content in these minerals is at the level of one part per million~\cite{Stewart1963MarineEvaporites}.

To overcome this obstacle, we target inelastic DM using suitable radiopure minerals containing heavy elements, while accounting for the history of the galactic DM velocity tail. The core ideas remain unchanged: we continue to have the same requirements on SSTD materials and the same read-out methods; the background mitigation strategies and the three irreducible backgrounds discussed in Section~\ref{sec:conventional_paleodectors} also remain the same. Nevertheless, we find that some aspects of the conventional paleodetectors described in Section~\ref{sec:conventional_paleodectors} are modified, including the mineral's chemical makeup, geological source, preferred mineral age, and requirements on backgrounds, depth, and exposure. We describe these novel aspects in detail below.

\subsection{Lead-bearing minerals}

First and foremost, the desired mineral must contain a large mass fraction of a heavy element. Of course, the heaviest naturally occurring element would be uranium itself ($A_{\rm U}=238$), but, as discussed in Section~\ref{sec:conventional_paleodectors}, its radioactivity would lead to a prohibitively large background; the same is true for thorium ($A_{\rm Th}=232$).

Instead, we find lead ($A_{\rm Pb}=207$) to be a good alternative. Naturally occurring lead is dominated by observationally stable isotopes~\cite{de_marcillac_experimental_2003,magic_number_Mayer_1948}. Although radioactive isotopes such as $^{210}\text{Pb}$ can occur as trace contaminants, they are all part of the decay chains of unstable elements such as $^{238}\text{U}$. So long as the mineral's geological source is radiopure, the mineral would not have inherited these radioactive isotopes.

Bismuth ($A_{\rm Bi}=209$) is another good candidate, with the naturally occurring $^{209}\text{Bi}$ being effectively stable for all practical purposes~\cite{de_marcillac_experimental_2003}, though it is usually rarer than Pb~\cite{Geothermal_fluid_NewZealand_2023}. In this work, we focus on Pb-bearing minerals, and leave the exploration of other heavy elements for future work.

As a concrete example, we focus on the mineral Laurionite (PbClOH)~\cite{ralph_mindatorg_2025,MindatLaurionite}, a colorless to white crystal, as a promising candidate for heavy-element paleodetectors, though we emphasize that there are many other potential mineral candidates (see Appendix~\ref{sec:other_minerals}). Laurionite has a resistivity $\rho_r \approx 10^{8}\text{--}10^{10}~\Omega~\text{cm}$~\cite{laurionite_conductivity}, comfortably satisfying \eq{\ref{eq:resistivity_requirement}}, the condition for materials to record damage tracks. Laurionite contains Pb with a mass fraction of $w_{\rm Pb}=0.8$. Furthermore, it also contains H, which drastically reduces fast neutron backgrounds (see Section~\ref{sec:irreducible_background}). As a chloride, it is soluble in water~\cite{MindatLaurionite}, a property relevant for its geological formation, which we turn to next. 

Candidate minerals must have viable geological formation mechanisms consistent with high concentrations of heavy elements, radiopurity, sufficient underground depth, and survival over timescales of tens of millions to billions of years. While our result is agnostic to the precise formation mechanism so long as the above criteria are satisfied, we identify a particular example below that we term ``brine precipitates (BPs)." This example is distinct from MEs and UBRs that have been studied in the context of paleodetectors thus far, which, as discussed above, typically do not contain heavy elements.

The starting point is geothermal brine, which refers to hot groundwater containing high concentrations of dissolved salts due to its long-term interaction with surrounding rocks. It is often geological in age, and typically found deep underground,  with temperatures $T \gtrsim 100\,^\circ\mathrm{C}$ at $\sim 5~\text{km}$ depth~\cite{Regenspurg2010}. It usually contains high concentrations of chlorine (Cl), sodium (Na), and calcium (Ca)~\cite{Regenspurg2010,KharakaHanor2003}, and often also high concentrations of heavy metals, including Pb~\cite{regenspurg_formation_2016,KharakaHanor2003,Regenspurg2010,Regenspurg2014,McKibben1987SaltonSea,Scheiber2013ScalingInhibitorSoultz}.
When the brine becomes oversaturated with respect to a given solid compound, the system becomes unstable to precipitation, which can be triggered by even slight changes in pressure, temperature, pH, or redox conditions, resulting in scale formation, i.e., the deposition of solid minerals on the surrounding rock surfaces~\cite{Regenspurg2010,brine_precipitate_review_2025}. These minerals are the BPs considered in this work.

The aforementioned mineral Laurionite (PbClOH) could precipitate from brines with sufficiently high concentrations of Pb and Cl, and with pH between about 6.4 and 10~\cite{regenspurg_formation_2016}\footnote{In fact, BP formation of Laurionite could be such a significant effect that it contributed to clogging up the geothermal well at Gro{\ss} Sch{\"o}nebeck in Germany~\cite{regenspurg_formation_2016,laurionite_clogging}.}. Another common Pb-bearing precipitate is Galena (PbS)~\cite{galena_and_laurionite_precipitate_2021}. Although Galena appears unsuitable, since its electrical resistivity~\cite{mineral_resistivity_1950} fails to satisfy the requirement in \eq{\ref{eq:resistivity_requirement}}, Ref.~\cite{paleodetectors_refine_2025} suggests otherwise; we therefore leave a detailed investigation of Galena to future work. We also briefly comment on other Pb-bearing minerals in Appendix~\ref{sec:other_minerals}.

Geothermal brine can have extremely low uranium concentrations. In the most radiopure cases to our knowledge, samples from geopressured, geothermal aquifers in the Gulf of Mexico basin were found to have uranium concentrations as low as\footnote{In fact, this uranium concentration should be considered an upper limit since the measurement technique likely overestimated~\cite{Kraemer_lowU_1981}.} $C^{238} \lesssim 3\times 10^{-12}~\gpg$~\cite{Kraemer_lowU_1981,Kraemer_lowU_1986,geopressure_review_1992}, lower than even the optimistic value of MEs in existing paleodetector literature\footnote{Some less radiopure but still promising examples elsewhere in the world: the Kawerau geothermal well in New Zealand reported $C^{238} =10^{-10}~\gpg$~\cite{Geothermal_fluid_NewZealand_2023}; the Bruchsal geothermal site in Germany found the radioactivity of U and Th to be below detection sensitivity, setting an upper limit $C^{238} <4 \times 10^{-10}~\gpg$~\cite{kolbel_waterrock_2020}.}~\cite{paleodetectors_details_2018,Baum:2019fqm}. See Appendix~\ref{sec: mineral_locations} for more details on these promising sources.

The cause of the low uranium concentrations is the reducing conditions of the environment\footnote{In the case of the U.S. Gulf Coast aquifers, reducing conditions are caused by organic matter in the environment during aquifer formation, and persist to the present day~\cite{Kraemer_lowU_1981}.}. 
Under reducing conditions, uranium takes the form of $\text{UO}_2$ or $\text{USiO}_4$, which are highly insoluble~\cite{Kraemer_lowU_1981,Kraemer_lowU_1986,kolbel_waterrock_2020,geochemistry_Krauskopf_1967,lowU_calculation_Langmuir_1978,uranium_solubility_Goodwin_1980}; in contrast, under oxidizing conditions, uranium forms $\text{UO}_2^{2+}$ or $\text{UO}_2(\text{CO}_3)_2^{2-}$, which are soluble~\cite{geochemistry_Krauskopf_1967}. In fact, calculations show that at an optimal redox potential and pH, uranium concentration could get as low as $C^{238} \lesssim 10^{-13}~\gpg$~\cite{lowU_calculation_Langmuir_1978,Kraemer_lowU_1981,uranium_solubility_Goodwin_1980}. Ref.~\cite{kolbel_waterrock_2020} argues that thorium (Th) is similarly insoluble. These considerations of the physical origin of radiopurity could aid the search for new geological sources of paleodetectors.

Precipitates from these radiopure brines are expected to inherit the extremely low uranium concentrations~\cite{Kraemer_lowU_1981}. We will take as benchmark the lowest measured concentration,
\begin{align}
    C^{238}_{\rm BP} = 3\times 10^{-12}~\gpg ~.
\end{align}
Unless otherwise specified, this benchmark concentration will be assumed in the rest of this paper. The minimum theoretical value from the aforementioned calculation,
\begin{align}
    C^{238}_{\rm BP,min} = 10^{-13}~\gpg ~,
\end{align}
will be considered an ultimate reach.

\subsection{Mineral age}
For conventional paleodetectors, exposure to DM is maximized by finding the oldest minerals, up to $\sim$~Gyr in age. However, in the context of inelastic dark matter, we find that the targeted astrophysical history of the DM velocity tail does not necessarily favor the oldest rocks.

As reviewed in Section~\ref{sec:LMC}, the fastest local DM may have originated from the LMC, which had its pericenter approach to the MW about $\sim 50$~Myr ago. For $\delta \gtrsim 550$~keV, the Higgsino could not upscatter without this LMC velocity boost. In this case, a Gyr-old mineral would only be accumulating backgrounds during the first 950~Myr, and would only start recording signal in the last $50$~Myr. In contrast, a much younger $50$~Myr-old mineral would have a better signal-to-noise ratio, if this speculation about the history of the DM velocity distribution is correct.

We emphasize that, for generic inelastic DM parameter space, we are interested in probing both smaller cross sections and larger mass splittings (see the two axes of Fig.~\ref{fig: projection}), and different mineral ages could cover different parts of parameter space. Probing smaller cross sections at moderate mass splittings $\delta \lesssim 500$~keV is insensitive to the velocity tail and benefits from older minerals, whereas probing larger mass splittings (e.g. for the Higgsino) is sensitive to the history of the velocity tail and would favor a 50~Myr-old sample.

This relatively young optimal age could make suitable minerals easier to find. For example, some of the aforementioned U.S. Gulf Coast aquifers with the lowest known uranium concentrations happen to be from the Oligocene age (23--34 Myr ago)~\cite{Kraemer_lowU_1981,Kraemer_lowU_1986}. Another benefit of the relaxation of the age requirement is a proportionally shorter required annealing time.

Note that even though $\Trock \sim 50$~Myr is much younger than the usual $\sim 1$~Gyr assumption in conventional paleodetectors, \eq{\ref{eq:1alpha_number}} remains valid because $\Trock$ is still much older than the half-life of $^{234}\text{U}$. 

Due to the uncertainty in astrophysical history, we also consider the purely SHM scenario, in which case a Gyr-old mineral remains preferable.

\subsection{Large signal and accessible samples}

An obstacle facing the paleodetector program is the difficulty of obtaining radiopure samples from 5-km depth. Such a low-background sample is necessary to probe new parameter space for elastic WIMPs.

The situation is different for inelastic electroweak WIMPs. The target Z-mediated cross section $\sim 10^{-39}~\text{cm}^2$ is many orders of magnitude larger than the elastic cross section probed in conventional paleodetectors. Consequently, as we will show in Section~\ref{sec: DM_signal}, for the lowest viable $\delta$, the DM signal is much larger than the backgrounds. There is thus an opportunity to explore new parameter space even with suboptimal (but much more accessible) mineral samples, that is, samples with larger radioactive and cosmogenic backgrounds, or less exposure.

Indeed, we demonstrate in Section~\ref{sec: projection_larger_bg} that lead-bearing samples with uranium concentrations on the order of one part per million by weight and cosmogenic backgrounds at the level of samples obtained from $\sim 2$~km depth can still probe new Higgsino parameter space. This opens up vastly more accessible sample sources, since uranium concentrations at this level are comparable to those of ordinary minerals in the Earth's crust, and the required cosmogenic backgrounds correspond to samples from depths comparable to those of existing underground laboratories. This exciting possibility provides an immediately actionable item for experimentalists. We also show in Section~\ref{sec: projection_smaller_exposure} that even samples with target volumes a million times smaller than usually required can still probe deep into unconstrained parameter space.

Of course, it is always desirable to have less background to probe the largest $\delta$ possible. As mentioned above, achieving the largest possible reach still requires 5-km depth and carefully chosen radiopure samples. Unless otherwise specified, we always refer to the optimal sample in this work.

\section{Dark matter signal}
\label{sec: DM_signal}

\subsection{Signal rate}

\begin{figure*}[t]
    \centering
    \begin{subfigure}{0.49\textwidth}
        \centering
        \includegraphics[width=\textwidth]{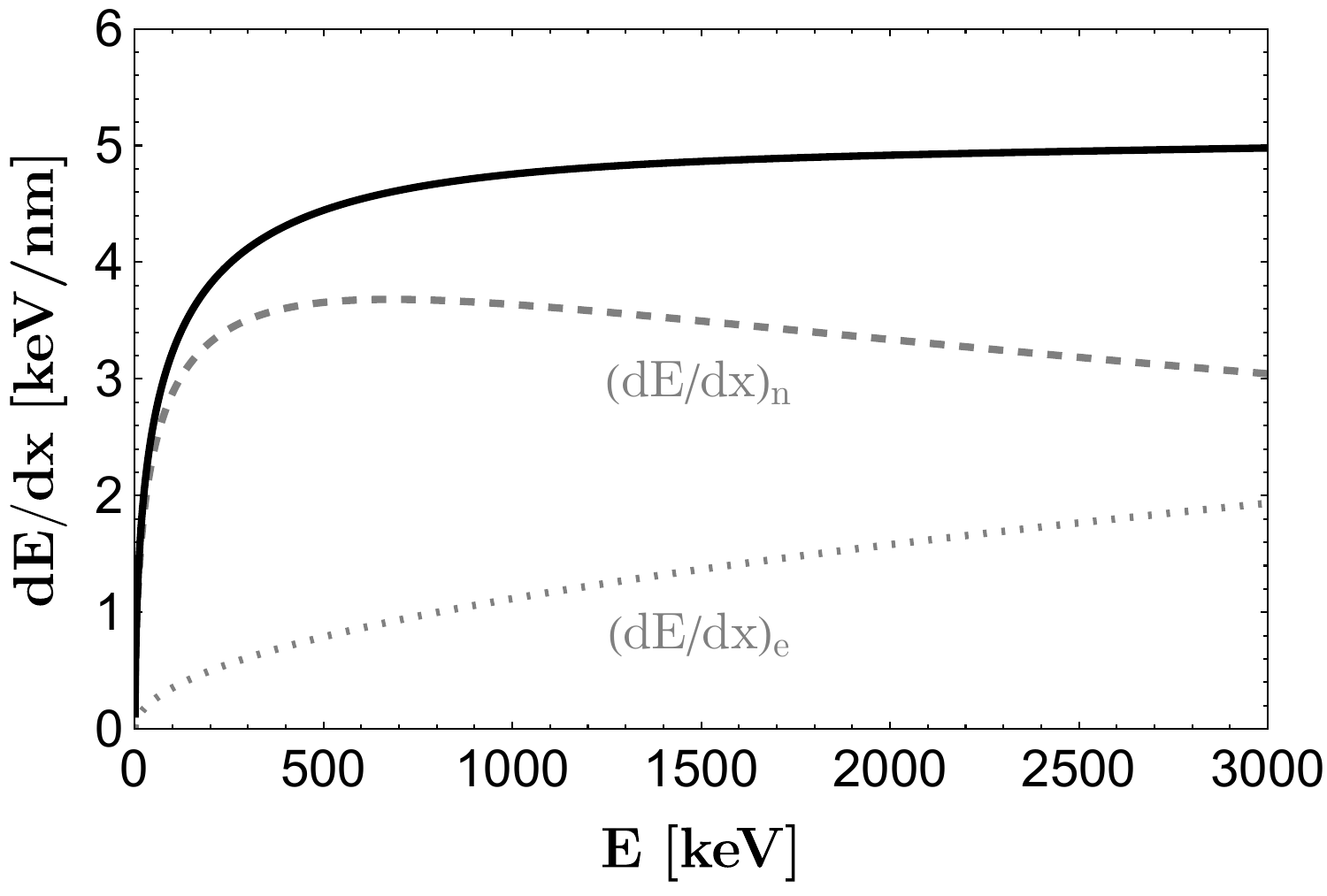}
        \caption{}
        \label{fig: dEdx}
    \end{subfigure}
    \hfill
    \begin{subfigure}{0.49\textwidth}
        \centering
        \includegraphics[width=\textwidth]{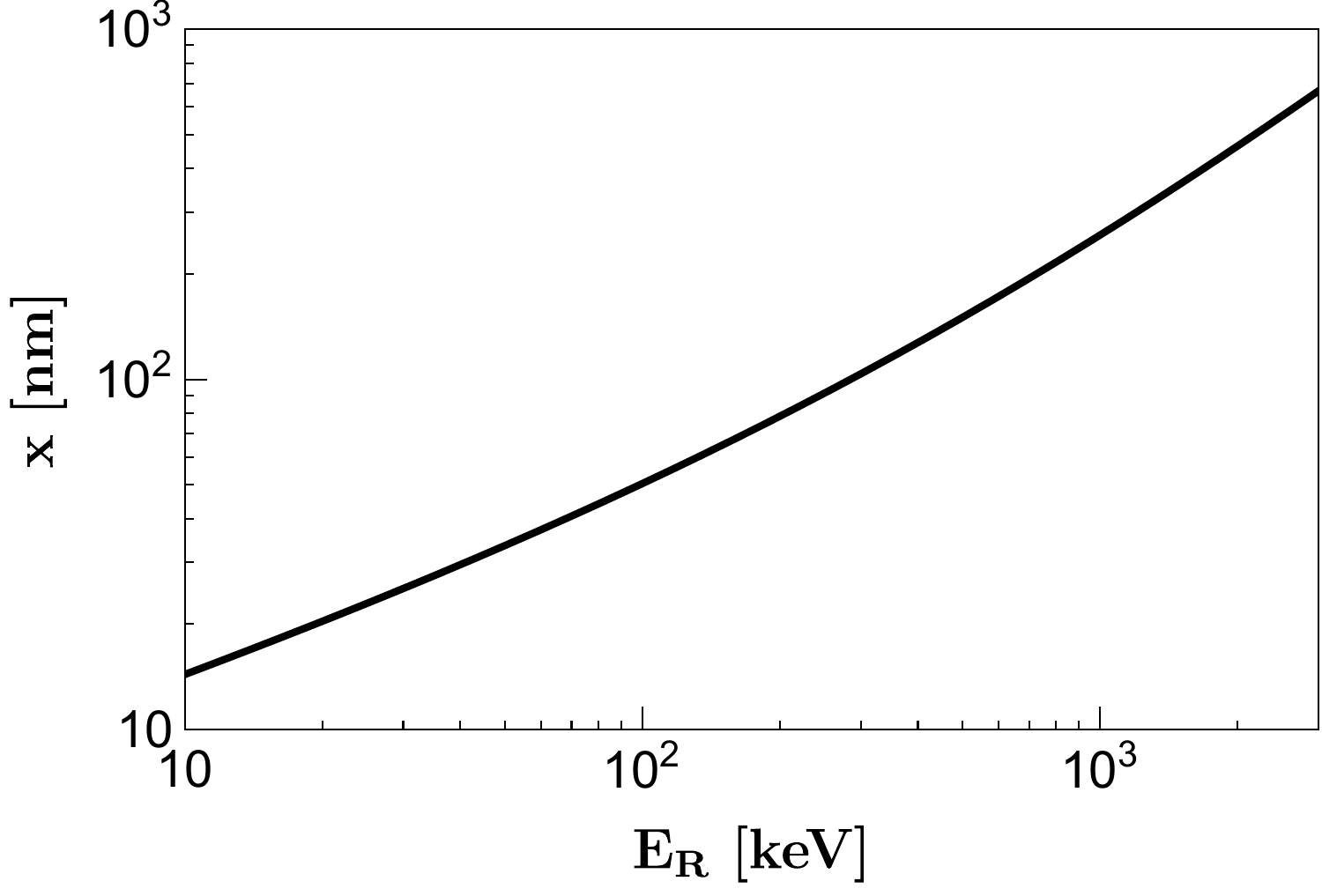}
        \caption{}
        \label{fig: x vs ER}
    \end{subfigure}
    
    \caption{For a scattered Pb nucleus in Laurionite (PbClOH): (a) the stopping power $dE/dx$, with both the nuclear and electronic contributions shown; (b) track length $x$ as a function of recoil energy $E_R$.}
    \label{fig: Pb_in_Laurionite}
\end{figure*}

We now compute the DM signal in the mineral track-length spectrum. The differential DM-nucleus scattering rate with respect to recoil energy is given by~\cite{EnhancingHiggsinoDM_2024}
\begin{align} \label{eq: rate ER}
    \frac{d \GxA}{dE_R} (E_R) &= \frac{m_A A^4 \nx  \sxn}{2\muA^2} ~\eta(\vmin(E_R)) ~F_A^2(E_R) ~,
\end{align}
where $\nx=\rhoDM/\mx$ is the DM number density with $\rhoDM = 0.3~\text{GeV}/\text{cm}^3$, $\sxn$ is the DM-nucleon cross section, $\vmin(E_R)$ is the minimum velocity~[\eq{\ref{eq: vmin}}], and we ignore the possible time dependence in the halo integral $\eta$~[\eq{\ref{eq:halo integral}}] for now. 
Coherent scattering is reflected in the $A^4$ factor~\cite{Barn}, and its correction due to the finite size of the nucleus is accounted for by the Helm form factor~\cite{Math_of_DM_Lewin_Smith_1995},
\begin{align} \label{eq: Helm form factor}
    F_A(E_R) = \frac{3}{q r_n} J_1(q r_n) e^{-q^2 s^2/2}~,
\end{align}
where $J_1(x)$ is a spherical Bessel function, $q=\sqrt{2m_A E_R}$ is the momentum transfer, $s \approx 0.9$~fm, and $r_n \approx 1.14 \left(A/0.93\right)^{1/3}$~fm.

Since the observables are the tracks in the mineral created by the recoiling nuclei, we relate $E_R$ to track length $x$ via the stopping power $dE/dx$. We use the semi-analytic treatment of $dE/dx$ described in Appendix~\ref{app: stopping power}; the result for a Pb nucleus in Laurionite (PbClOH) is shown in Fig.~\ref{fig: dEdx}.
The track length is given by the integral~\cite{paleodetectors_2018}
\begin{align} \label{eq: x(ER)}
    x(E_R) = \int_0^{E_R} dE' \left| \frac{dE}{dx} (E') \right|^{-1} ~.
\end{align}
In Fig.~\ref{fig: x vs ER}, we show the numerically integrated result for Pb in Laurionite.

Since $x(E_R)$ is a one-to-one function, we can numerically invert this to get $E_R(x)$, allowing us to use $x$ as the independent variable.
Taking the derivative of \eq{\ref{eq: x(ER)}} and substituting $E_R(x)$, we obtain the distribution of $E_R$ with respect to $x$,
\begin{align}
   \frac{dE_R}{dx}(x) \equiv \frac{1}{x'(E_R(x))} = \left| \frac{dE}{dx} (E_R (x)) \right|  ~.
\end{align}
We then obtain $d\GxA/dx(x)$ via the chain rule.

To get the DM track-length spectrum in a mineral sample, we multiply by the total number of target nuclei, 
\begin{align}
    \frac{d \Gxrock}{dx}= \left( \frac{w_A \mrock}{m_A} \right) \frac{d\GxA}{dx} \equiv w_A \rho V \frac{d\RxA}{dx} ~,
\end{align}
where $w_A$ is the mass fraction of $A$, assumed to be the only heavy element present\footnote{This is the case for Pb in Laurionite (PbClOH). Even at low $\delta$ where other elements can be scattered, the signal from Pb dominates due to the coherent enhancement in the scattering rate. Hence, we conservatively ignore signals from other elements.}; $\RxA$ is the scattering rate per target-nucleus mass; and $M$, $\rho$, and $V$ are the mass, density, and volume of the mineral sample, respectively. The processable volume is assumed to be $60~\text{cm}^3$ for SAXs and $6~\text{mm}^3$ for HIBM. The density $\rho$ is larger for heavy-element minerals, giving them an enhancement relative to lighter materials; e.g. Laurionite, with a density of $\rho=6.24~\text{g}/\text{cm}^3$~\cite{anthony_handbook_mineralogy}, is 3.7 times denser than Epsomite.

\subsection{True spectrum}

\begin{figure*}[t]
    \centering
    
    \begin{subfigure}{0.49\textwidth}
        \centering
        \includegraphics[width=\textwidth]{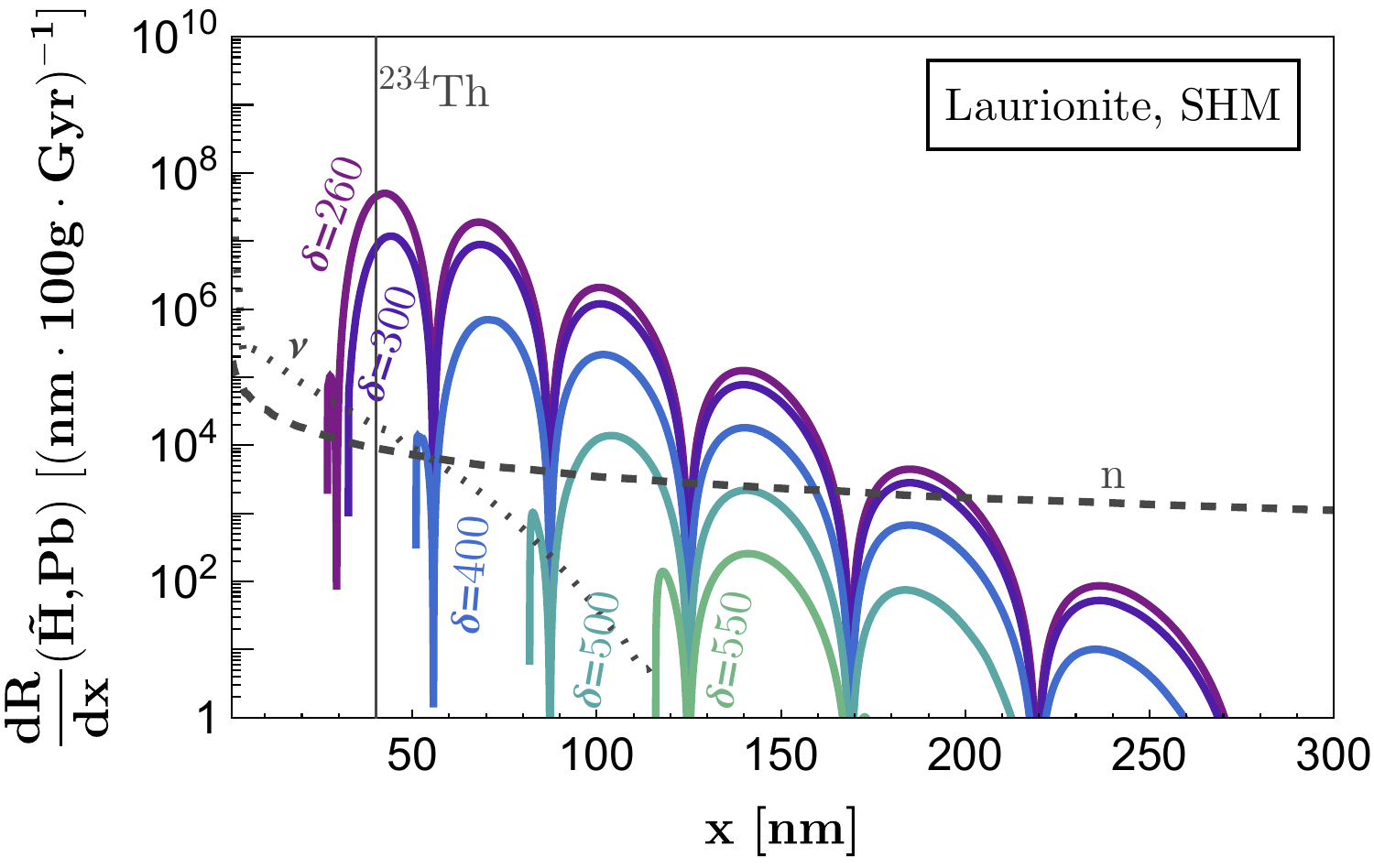}
        \caption{}
        \label{fig: spectra_SHM}
    \end{subfigure}
    \hfill
    \begin{subfigure}{0.49\textwidth}
        \centering
        \includegraphics[width=\textwidth]{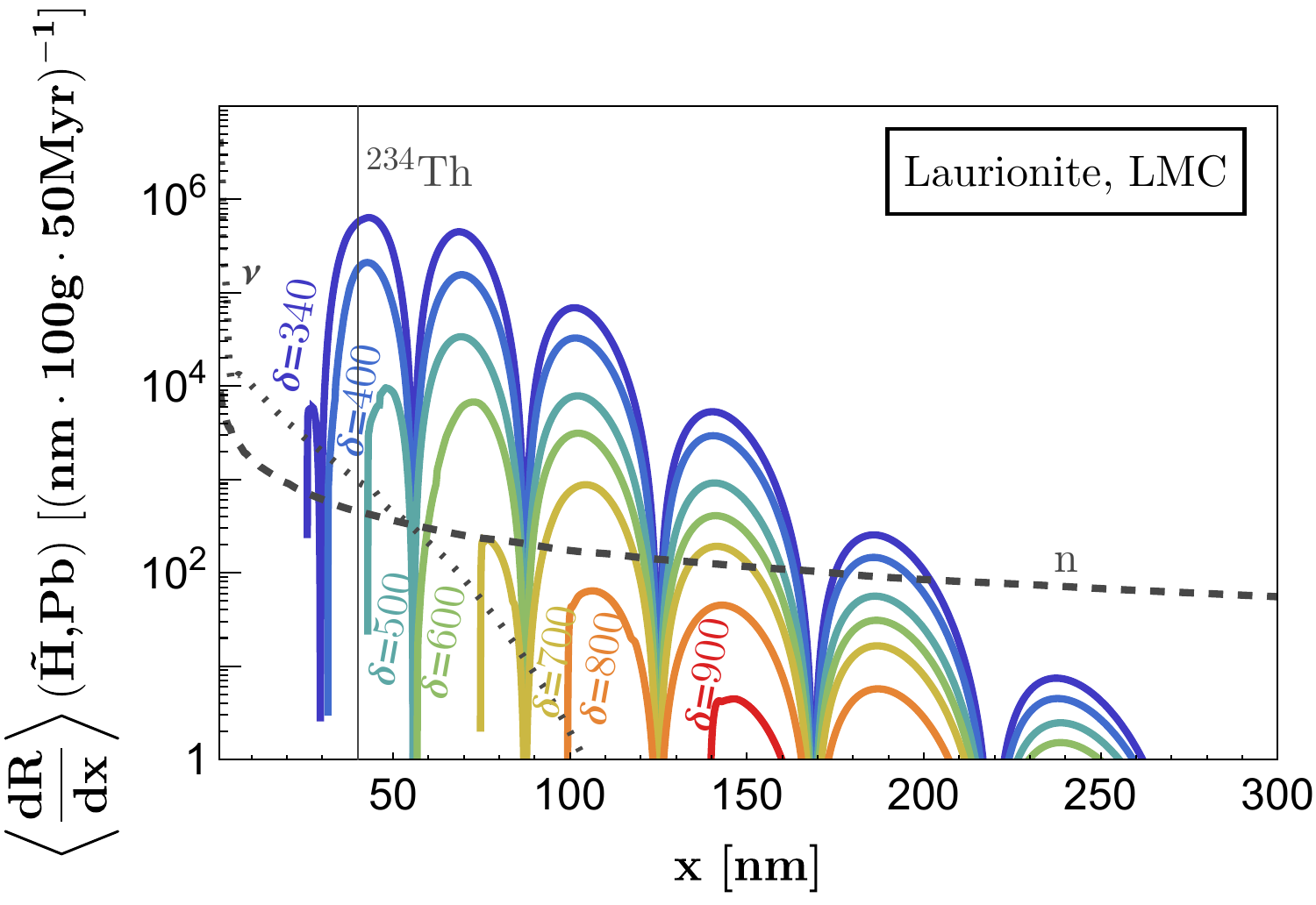}
        \caption{}
        \label{fig: spectra_LMC}
    \end{subfigure}
    
    \caption{Higgsino-induced track-length spectrum per target nucleus mass for Pb in Laurionite (PbClOH) for various mass splittings $\delta$ in units of keV, as well as the expected backgrounds from neutrinos, neutrons, and $1\alpha$ events ($^{238}\text{U} \to ^{234}\text{Th}+\alpha$), assuming (a) the Standard Halo Model (SHM); (b) the LMC model for a 50~Myr-old mineral, with the time-averaged spectrum shown.}
    \label{fig: true_spectra}
\end{figure*}

Fig.~\ref{fig: spectra_SHM} shows the Higgsino-induced track-length spectrum per target-nucleus mass, $dR_{\Ht, \rm Pb}/dx$, for Pb in Laurionite for various $\delta$, assuming the SHM. The largest possible signal at the current limit ($\delta=260$~keV) is more than 3 orders of magnitude larger than the background. As $\delta$ increases, the minimum velocity required for scattering increases; hence, the signal is increasingly suppressed as the allowed phase space is pushed further into the DM velocity tail. However, even at $\delta=500$~keV, the signal is above background.

In addition, larger $\delta$ narrows the allowed range of recoil energies $E_R$ while shifting the typical recoil energy to higher values, as seen in the contour plot of Fig.~\ref{fig: ER contour}. Because $x$ increases monotonically with $E_R$, the same behavior appears in the track-length spectrum: the average track length becomes longer as $\delta$ increases, while the allowed range of track lengths becomes narrower. Explicitly, for a given $\delta$, the minimum and maximum $x$ are given by $x(E_{R,\pm})$, where $E_{R,\pm}$ are the extremal recoil energies [\eq{\ref{eq: ER full euqation}} with $\cos^2 \theta=1$] at the maximum available velocity
\begin{align}
    E_{R,\pm} &= \frac{\mu_A}{m_A} \left[(\mu_A \vmax^2 - \delta )  
    \pm \sqrt{(\mu_A \vmax^2 ) (\mu_A \vmax^2 - 2 \delta )} \right] ~.
\end{align}

The oscillatory features in the spectrum are due to the Helm form factor~[\eq{\ref{eq: Helm form factor}}]. Since the size of the nucleus scales as $r_n \propto A^{1/3}$, a Pb nucleus is large enough that the relevant momentum transfer is probing deep into the nucleus. Such features do not show up in conventional paleodetectors because lighter nuclei are much smaller.

For the expected backgrounds, we have assumed the benchmark $^{238}\text{U}$ concentration of $C^{238}_{\rm BP}= 3\times 10^{-12}~\gpg$. Note that due to the mono-energetic nature of the $\alpha$-decay, the 1$\alpha$ background is a delta function in this ``true spectrum" (with its track length computed for the recoiling $^{234}\mathrm{Th}$ nucleus in Laurionite), but will gain a finite width once we take into account experimental uncertainties in the ``smeared spectrum."

To probe the DM from the LMC, we must contend with some astrophysical uncertainties. Ref.~\cite{LMC} only studied the distribution at two snapshots in time: at pericenter and at present day. As the LMC approached the MW $\sim 50$~Myr ago, the DM velocity distribution must have been continuously increasing (and could have peaked at a different time and a faster velocity than at the pericenter), but we conservatively omit this contribution as the details are unknown. An ideal mineral for probing large $\delta$ would then be a $50$~Myr-old mineral, because much older minerals would only accumulate extra background but not signal since we conservatively assume there were no fast DM particles before then.

Furthermore, more precise simulations are needed to determine how the halo integral interpolates between the two snapshots. We make the simple assumption that $\eta(v_{\min},t)$ evolves linearly between $t_{\rm peri}$ and $t_{\rm pres}$. 
Since the scattering rate depends on time only through $\eta$, the time-averaged spectrum $\left \langle dR_{\Ht, \rm Pb}/dx \right \rangle$ is obtained by replacing
$\eta(v_{\min},t)$ with its time average $\left \langle \eta(v_{\min})\right \rangle$, which, for a linearly evolving history, is given by the midpoint,
\begin{align}
\langle \eta(v_{\min})\rangle 
= \frac{1}{2} \left[ \eta(v_{\min},t_{\rm peri})+\eta(v_{\min},t_{\rm pres}) \right] ~.
\end{align}

While these assumptions appear reasonable, we hope the astrophysical uncertainties can be reduced with further effort in simulating the DM velocity distribution, as is motivated by this work.

The time-averaged spectrum for Pb in a 50 Myr-old sample of Laurionite in the LMC model is shown in Fig.~\ref{fig: spectra_LMC}. The largest possible signal at the current limit for the LMC ($\delta=340$~keV) is 3 orders of magnitude larger than the backgrounds, and remains above background up to $\delta \gtrsim 700$~keV. 
It can be shown that a more conservative assumption of the present-day LMC velocity distribution for the last 50~Myr would give a similar spectrum, but the sensitivity cuts off at somewhat smaller $\delta$.

\subsection{Smeared spectrum}

\begin{figure*}[t]
    \centering
    \begin{subfigure}{0.49\textwidth}
        \centering
        \includegraphics[width=\textwidth]{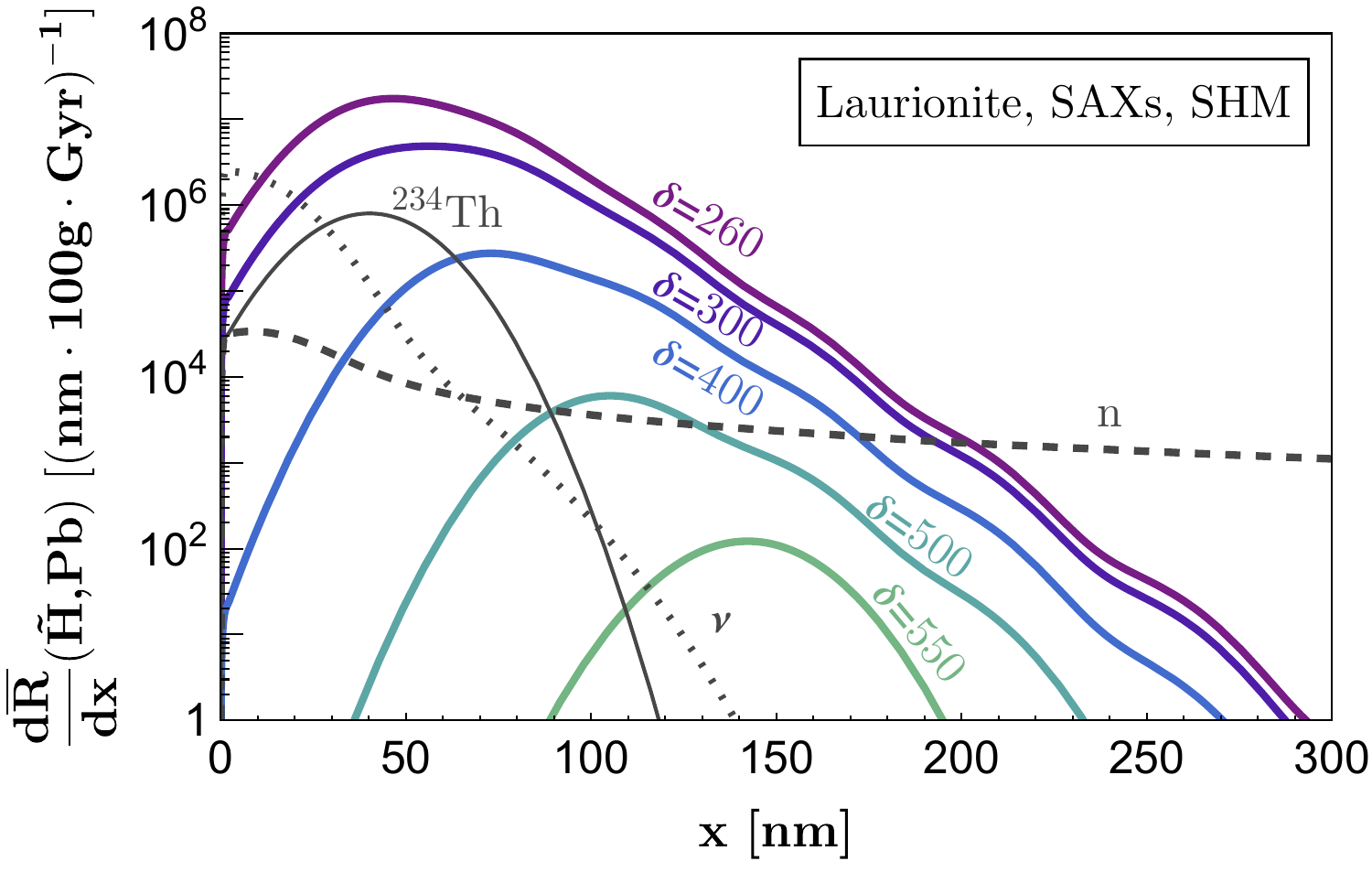}
        \caption{}
        \label{fig:obs_spectra_SHM}
    \end{subfigure}
    \hfill
    \begin{subfigure}{0.49\textwidth}
        \centering
        \includegraphics[width=\textwidth]{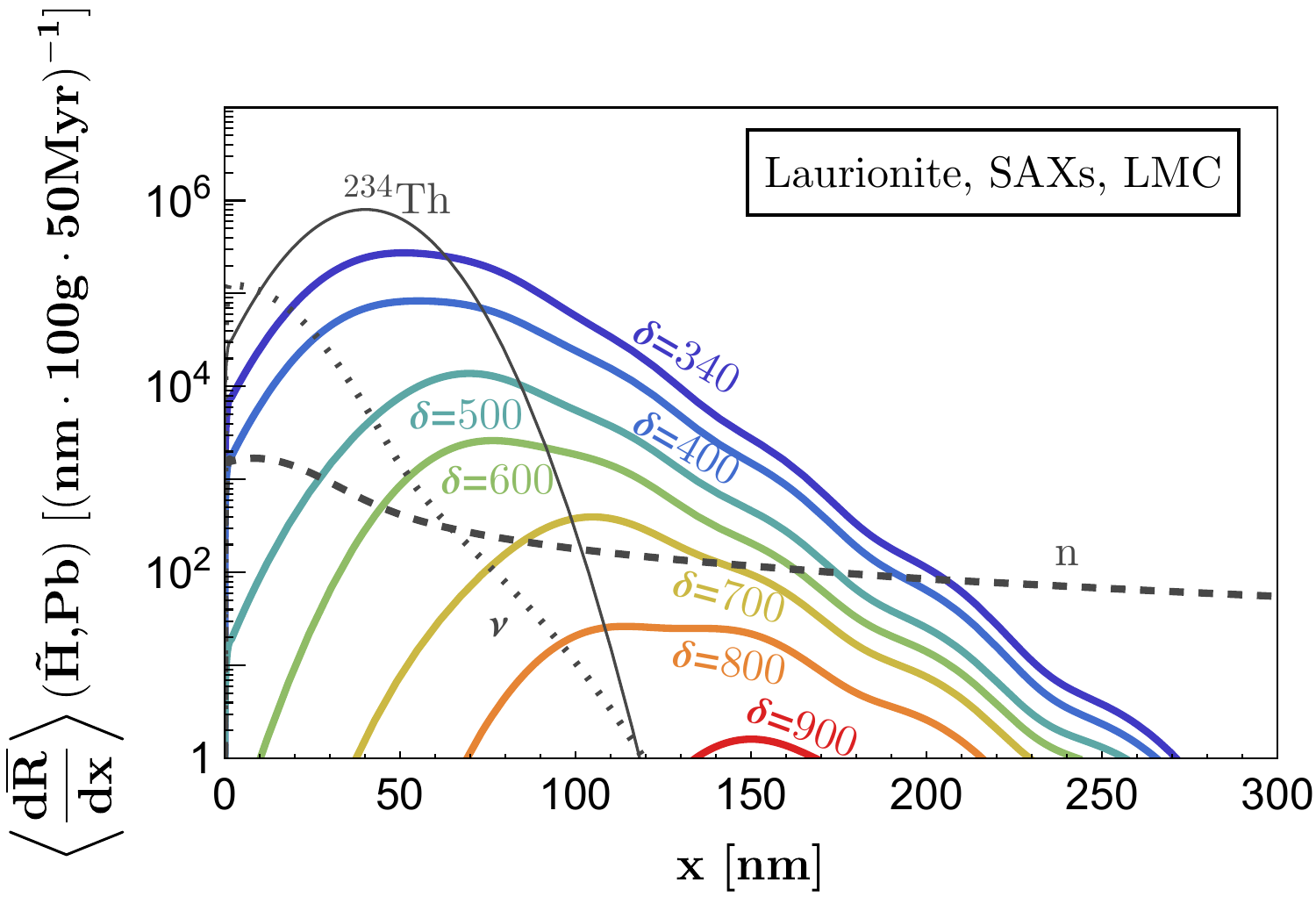}
        \caption{}
        \label{fig:obs_spectra_LMC}
    \end{subfigure}
    
    \caption{Same as Fig.~\ref{fig: true_spectra}, but after smearing against a truncated Gaussian with standard deviation $\sigma_x=15$~nm, corresponding to the uncertainty of the Small Angle X-ray scattering (SAXs) read-out method.}
    \label{fig: obs_spectra}
\end{figure*}

\begin{figure*}[t]
    \centering
    
    \begin{subfigure}{0.49\textwidth}
        \centering
        \includegraphics[width=\textwidth]{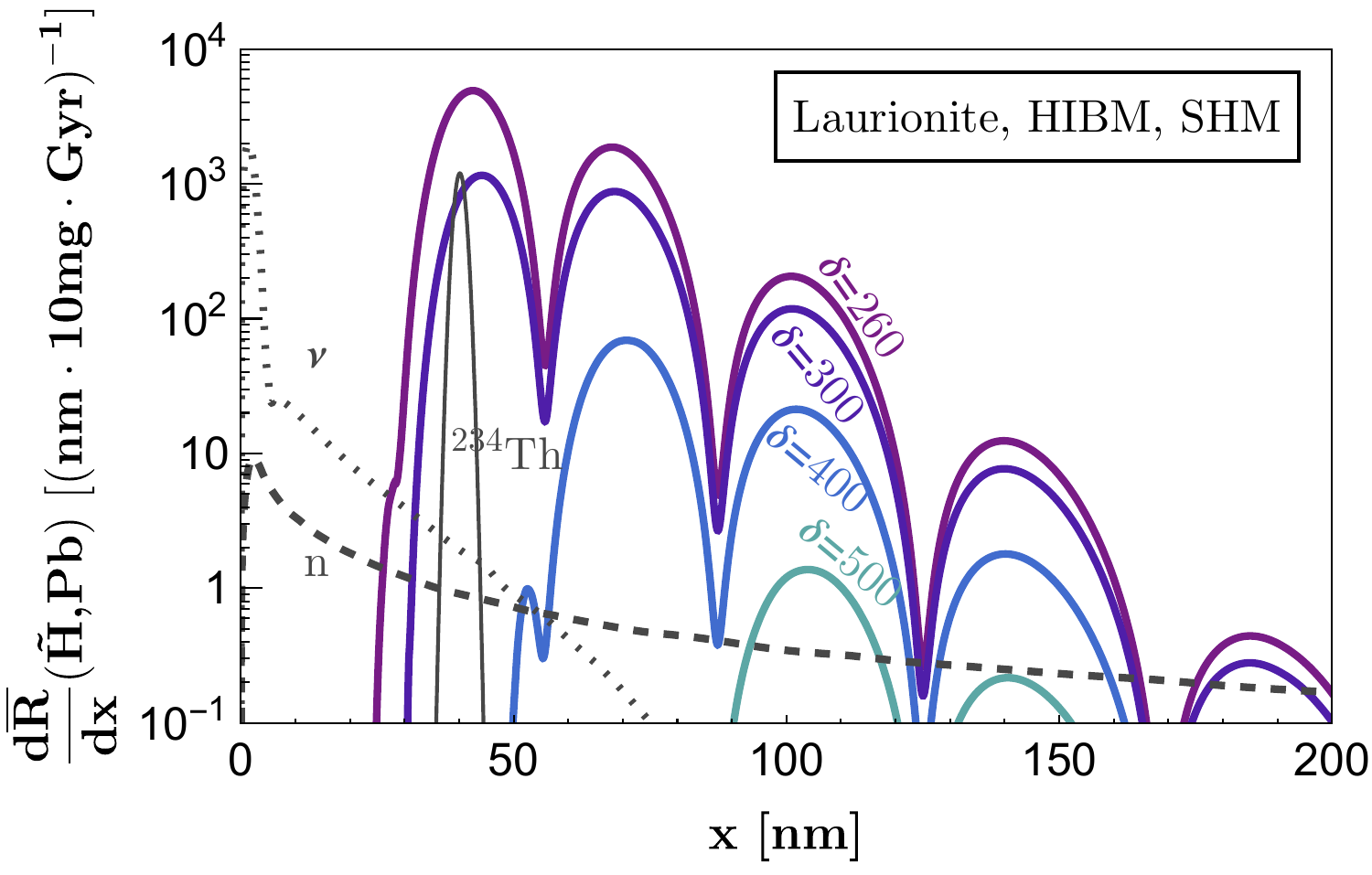}
        \caption{}
        \label{fig: obs_spectra_SHM_HIBM}
    \end{subfigure}
    \hfill
    \begin{subfigure}{0.49\textwidth}
        \centering
        \includegraphics[width=\textwidth]{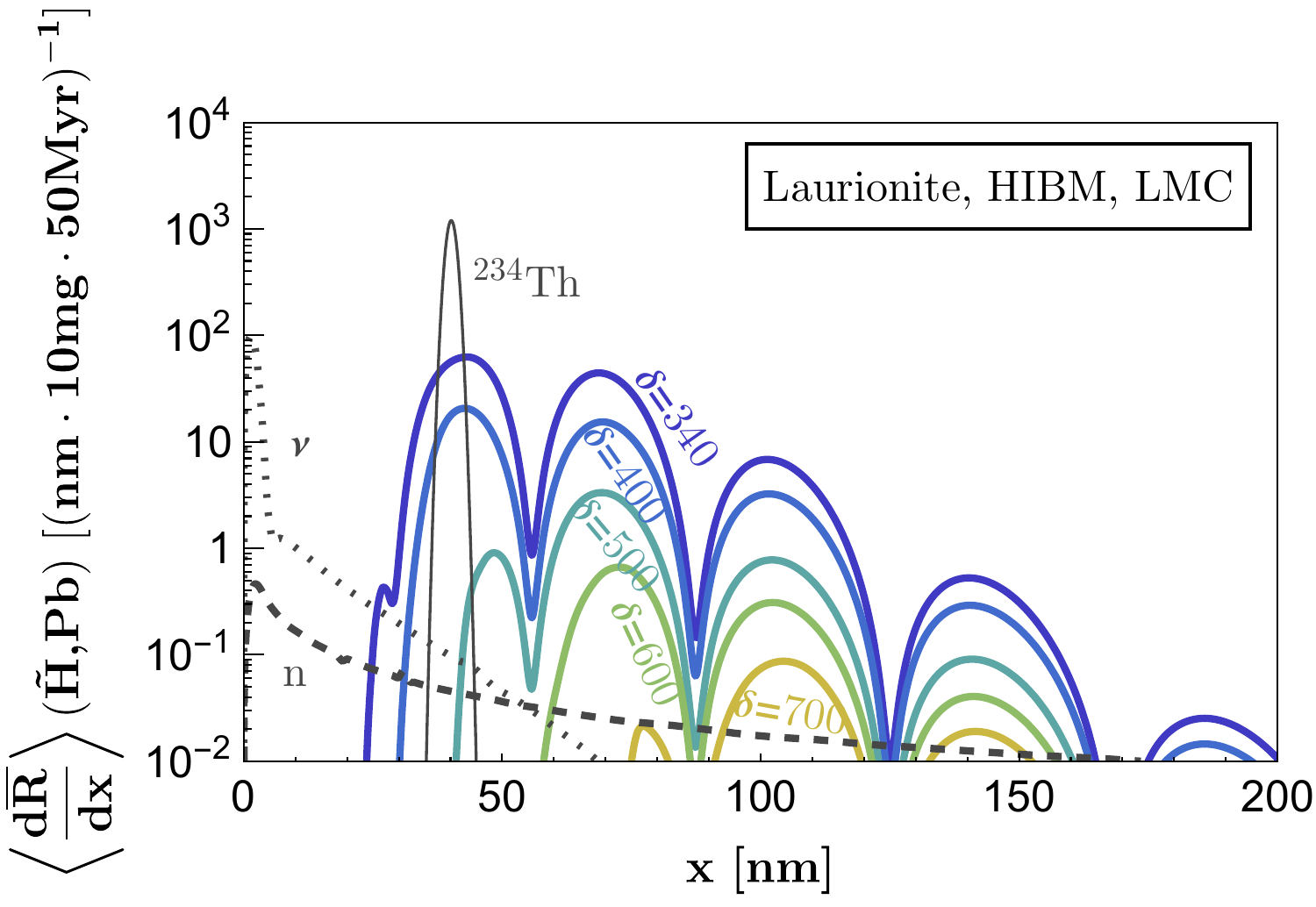}
        \caption{}
        \label{fig: obs_spectra_LMC_HIBM}
    \end{subfigure}
    
    \caption{Same as Fig.~\ref{fig: true_spectra}, but after smearing against a truncated Gaussian with standard deviation $\sigma_x=1$~nm, corresponding to the uncertainty of the Helium Ion Beam Microscopy (HIBM) read-out method.}
    \label{fig: obs_spectra_HIBM}
\end{figure*}

The measurement of the track length necessarily introduces uncertainty. We model the observed smeared spectrum as the convolution between the true spectrum and a normalized truncated Gaussian $G$ with standard deviation $\sigma_x$ set by the resolution of the read-out method:
\begin{align}
    \frac{d\overline{R}_{\chi A}}{dx}(x) \equiv \int_0^{\infty} dx' \frac{dR_{\chi A}}{dx'}(x')G(x-x',\sigma_x) ~,
\end{align}
where the bar denotes the smeared spectrum.
Recall from Section~\ref{sec: readout} that SAXs has a resolution of $\sigma_x=15$~nm, and HIBM has a resolution of $\sigma_x=1$~nm.

In Fig.~\ref{fig: obs_spectra}, we show the smeared spectrum for the SAXs method. 
The oscillatory features in the signal's true spectrum due to the Helm form factor are mostly washed out by the smearing. Note that while the true spectrum of the $1\alpha$ background is a delta function, smearing broadens it to a finite width. Because the 1$\alpha$ background is effectively time-independent, it breaks the age-target-mass degeneracy and has a larger impact in younger minerals.

We also show the smeared spectrum for the HIBM method in Fig.~\ref{fig: obs_spectra_HIBM}. In contrast to the SAXs method, the much smaller resolution preserves the oscillatory features. Although this method requires a $10^4$-fold reduction in target volume relative to the SAXs method, a mineral with exposure $60~\text{cm}^3~\text{Gyr}$ in the SHM is not exposure limited to begin with, except for the very largest $\delta$. Hence, Fig.~\ref{fig: obs_spectra_SHM_HIBM} shows a promising, low-background spectrum for a large range of track lengths, up to $\delta \sim 500$~keV. Similar statements apply to the LMC, though the younger mineral considered in this case is more exposure limited, and the sensitivity cuts off earlier at $\sim 700$~keV relative to the SAXs method (Fig.~\ref{fig: obs_spectra_LMC_HIBM}).

Of course, to probe the largest $\delta$ possible, the exposure will necessarily be a limit as the available phase space is pushed into the end of the DM velocity tail. So we consider the SAXs method when setting the ultimate projection. However, as we will show later, if we find a sample with much larger background, the signal-to-noise ratio could favor the HIBM method.

\section{Results and Projections}
\label{sec: projection}

We now calculate the projected limits. Define the smeared signal spectrum as
\begin{align}
    s(x) &\equiv \int_{0}^{\Trock} dt \, \frac{d \overline{\Gamma}_{\chi,\text{mineral}}}{dx}(x,t) \\
    &= V \rho w_A \int_{0}^{\Trock} dt \,  \frac{d \overline{R}_{\chi,A}}{dx}(x,t) ~,
\end{align}
where $T$ and $V$ are the age and volume of the mineral sample, respectively, and the time integral is nontrivial for the case of the LMC model. The smeared background spectra, $b_i(x)$ for $i = \nu,n,1\alpha$, are defined analogously\footnote{Except that the 1$\alpha$ background does not scale with time.}.
The total number of signal events is given by
\begin{align}
    S \equiv \int_0^{\infty}dx~ s(x) ~.
\end{align}

The optimal signal-to-noise ratio is found using the matched filter~\cite{matched_filter_1960}:
\begin{align}
    \SNR = \sqrt{\int_{0}^{\infty}dx~ \frac{s^2(x)}{\sum_{i} \left[ b_{i}(x)+\epsilon_{i}^2 b_{i}^2(x) \Delta x \right]}}~,
\end{align}
where $\Delta x=\sigma_x$ denotes the effective independent track-length interval, and the relative systematic errors are $\epsilon_{\nu}= 100\%$ and $\epsilon_{n}=\epsilon_{1\alpha}= 1\%$~\cite{paleodetectors_details_2018}. 
To set projected limits, we numerically determine, for each $\delta$, the smallest $\sxn$ that satisfies the conditions for 3 standard deviations and at least 5 signal events:
\begin{align}
    \SNR \geq 3,~ S \geq 5 ~.
\end{align}
Note that the large systematic errors for the neutrino background imply the sensitivity does not scale as the square root of the exposure.

All sensitivity plots in this work assume a dark matter mass of $1.1$~TeV, corresponding to the Higgsino thermal mass. It should be emphasized that these results can be extended to other masses and corresponding splittings, and hence this method has sensitivity to a broad range of models including other electroweak multiplets~\cite{Bottaro:2022one}, dark photon inelastic dark matter~\cite{Baryakhtar:2020rwy,Bramante:2016rdh}, and inelastic dark matter with dipole moments~\cite{MagneticInelasticDM,Baryakhtar:2020rwy,Bramante:2016rdh,Chang:2010en}. We leave a systematic analysis of this to future work. 

\subsection{Maximal reach}

\begin{figure}[t]
    \centering
    \includegraphics[width=0.5\textwidth]{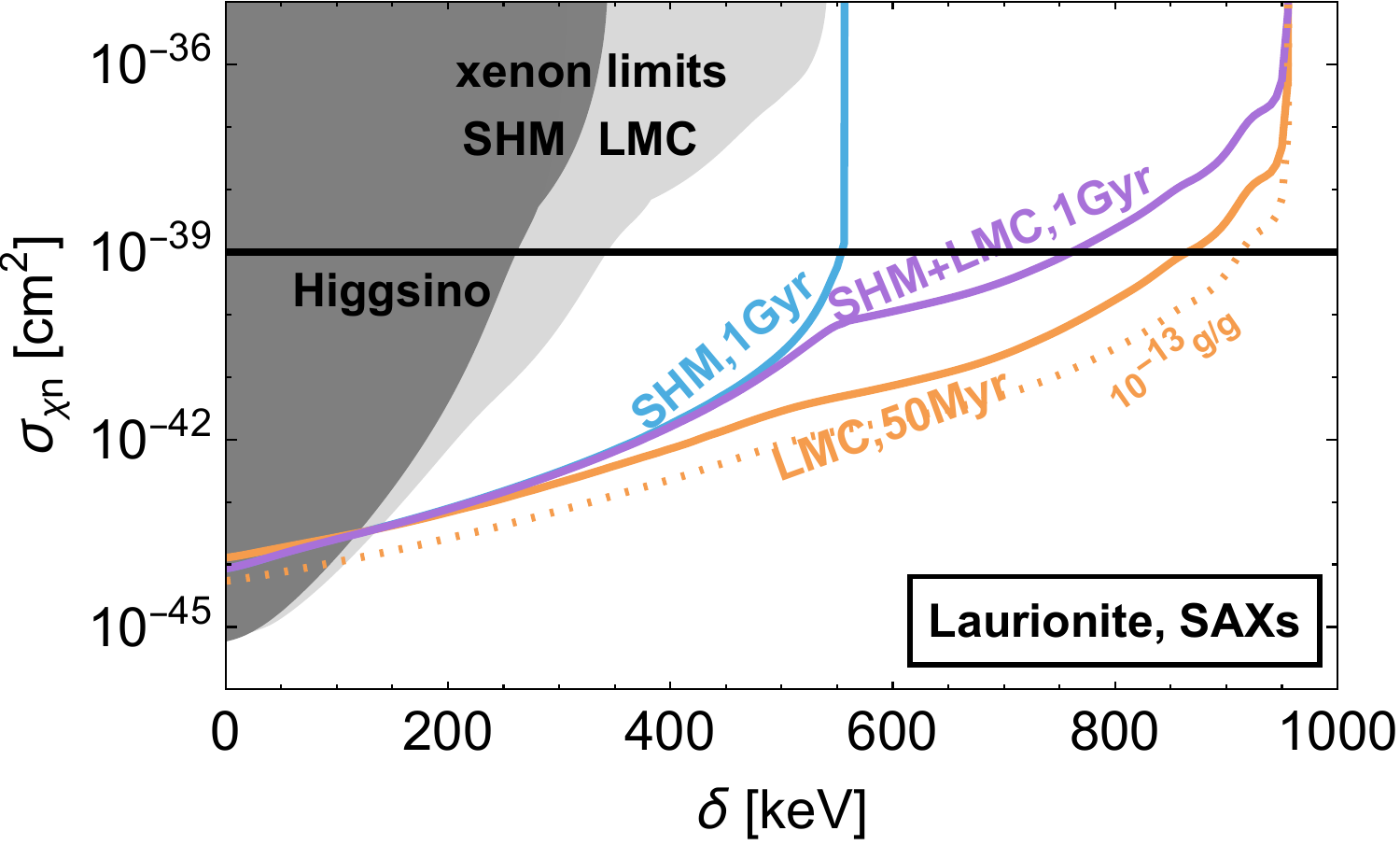}
    \caption{Projected limit on DM-nucleon cross section $\sxn$ vs mass splitting $\delta$, for a $60~\text{cm}^3$ sample of Laurionite (PbClOH) with resolution $\sigma_x=15$~nm relevant for Small Angle X-ray scattering (SAXs). Results are shown for a Gyr-old sample assuming the Standard Halo Model (SHM) (blue); the SHM followed by the LMC model during the last 50~Myr (purple); and a 50~Myr-old sample in the LMC model (orange). The orange dotted line assumes a theoretically minimal $^{238}\text{U}$ concentration of $C^{238} = 10^{-13}~\gpg$. The Higgsino cross section and existing limits~\cite{EnhancingHiggsinoDM_2024} from xenon experiments are also shown.
    }
    \label{fig: projection}
\end{figure}

\begin{figure*}[t]
    \centering
    \begin{subfigure}{0.49\textwidth}
        \centering
        \includegraphics[width=\textwidth]{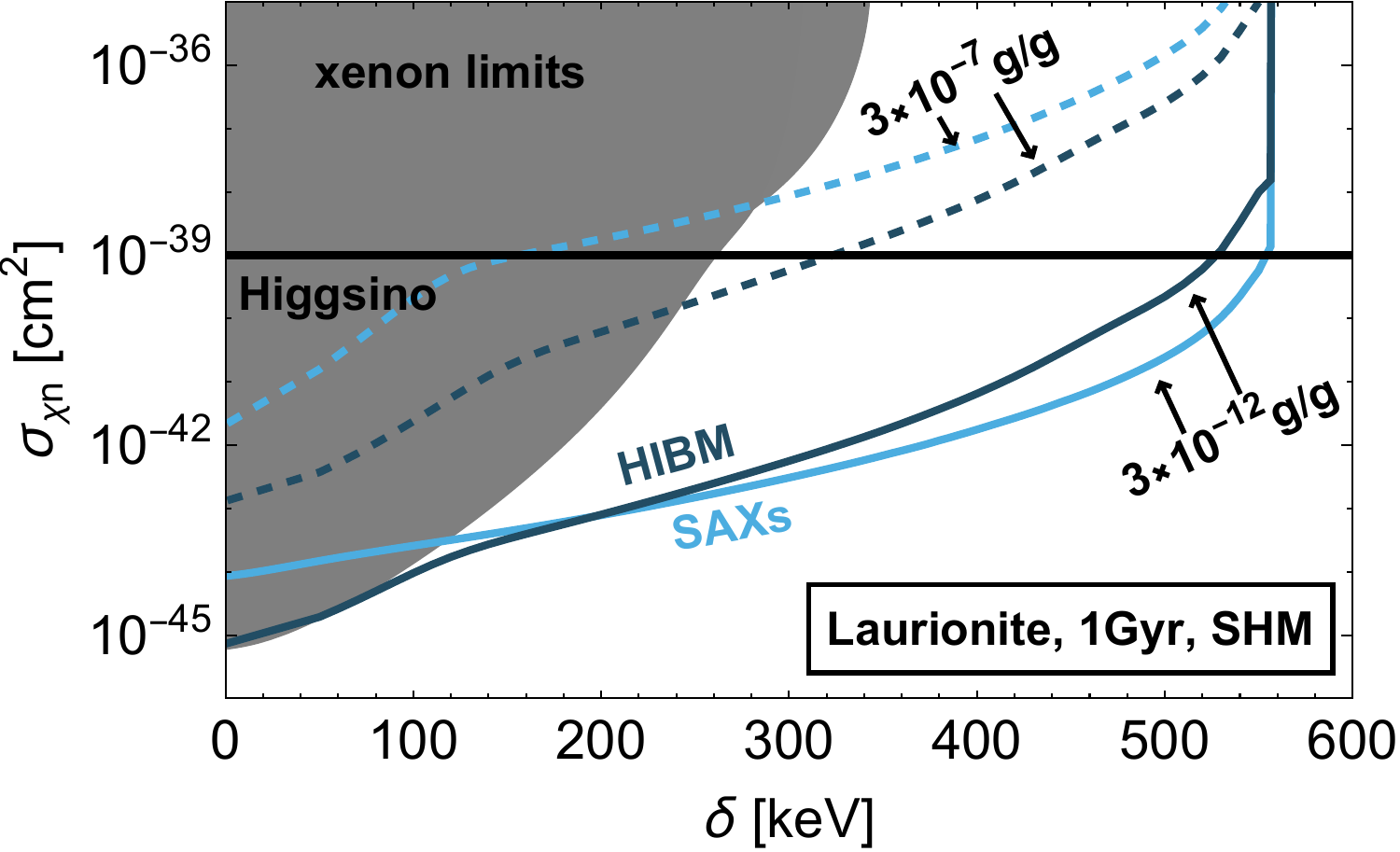}
        \caption{}
        \label{fig: SHM_larger_bg}
    \end{subfigure}
    \hfill
    \begin{subfigure}{0.49\textwidth}
        \centering
        \includegraphics[width=\textwidth]{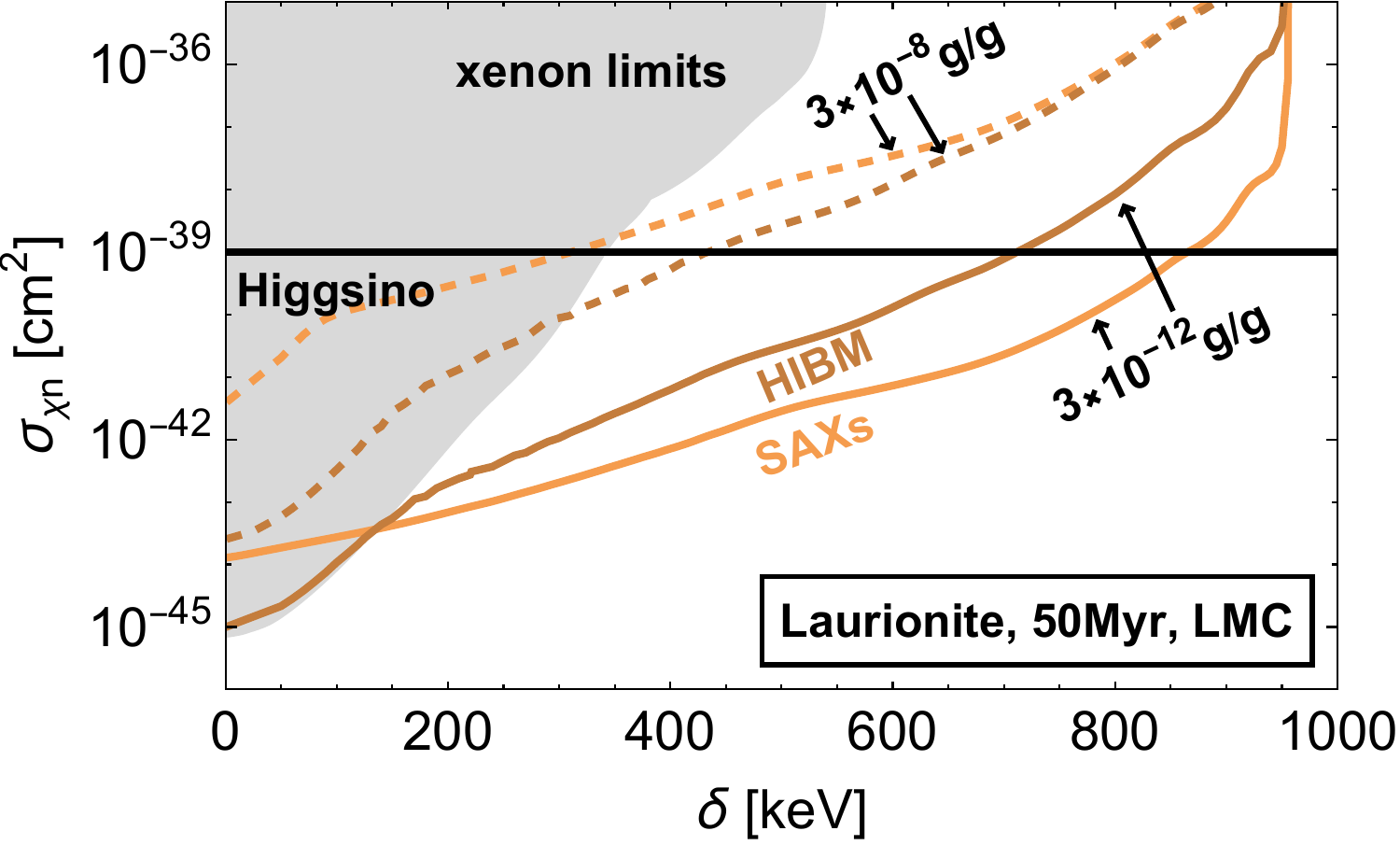}
        \caption{}
        \label{fig: LMC_larger_bg}
    \end{subfigure}
    
    \caption{Same as Fig.~\ref{fig: projection}, but we compare the read-out methods of Small Angle X-ray scattering (SAXs) ($60~\text{cm}^3$ sample, $\sigma_x=15~\text{nm}$) and Helium Ion Beam Microscopy (HIBM) ($6~\text{mm}^3$ sample, $\sigma_x=1~\text{nm}$). 
    Solid curves assume benchmark uranium concentration $C^{238}=3\times 10^{-12}~\gpg$. Dashed curves show interesting reach even for suboptimal samples with (a) $C^{238}=3\times 10^{-7}~\gpg$ for 1~Gyr; (b) $C^{238}=3\times 10^{-8}~\gpg$ for 50~Myr. 
    }
    \label{fig: projection_larger_bg}
\end{figure*}

The main result is shown in Fig.~\ref{fig: projection}, where we consider scanning a $60~\text{cm}^3$ sample of Laurionite with SAXs. In the SHM (blue curve), a Gyr-old sample extends the generic inelastic DM reach to splittings of $\approx 560$~keV, compared to the state-of-the-art $\approx 350$~keV with xenon. Furthermore, there is sensitivity down to   $10^{-39}\textrm{cm}^2$ cross sections for the largest splittings accessible to a Pb-based target.

The orange and purple curves correspond to also including fast dark matter originating from the recent LMC passage. For large enough cross section, this pushes the splitting sensitivity to more than 900 keV. In the rest of this section, we discuss these curves and other scenarios in the context of the Higgsino benchmark, which can be read off by the intersection of a given curve with the horizontal black line denoting the Higgsino target cross section.

Assuming the LMC model, a 50~Myr-old sample reaches $\delta\approx 870$~keV, shown by the orange solid curve. This splitting is even higher than that of a hypothetical present-day Pb detector with the same exposure ($\delta\approx840$~keV, not shown) owing to the speeds from the LMC being marginally higher 50 Myr ago.

We also make an ultimate LMC projection in which a BP sample is found with the theoretically minimal $^{238}\text{U}$ concentration, $C^{238}_{\rm BP,min}=10^{-13}~\gpg$ (whereas all other scenarios assume the experimentally measured value $C^{238}_{\rm BP}=3\times 10^{-12}~\gpg$). In this case, we find an ultimate reach of $\delta \approx 920$~keV, shown by the orange dotted curve.
This comes close to the largest possible splitting to which Pb is sensitive in the LMC, $\delta_{\rm max,Pb}^{\rm LMC}\approx 960~\text{keV}$.

Finally, we consider a Gyr-old mineral subject to the SHM for the first 950~Myr and additionally to the LMC for the last 50~Myr. As expected, the larger background weakens the reach of this older mineral relative to its younger counterpart, with a Higgsino limit of $\delta \approx 770$~keV, as shown by the purple curve.

\subsection{Larger backgrounds: 2~km depth and radio-impurity}
\label{sec: projection_larger_bg}

In Fig.~\ref{fig: projection_larger_bg}, we compare the reach of the SAXs and the HIBM read-out methods, accounting for their respective processable sample volumes and resolutions. In the SHM, for the benchmark background of $C^{238}=3\times 10^{-12}~\gpg$ and a 1~Gyr sample, the HIBM method has a reach of $\delta \approx 530$~keV, only 30~keV less than that of SAXs. In the LMC model with a 50~Myr sample, the HIBM method reaches $\delta \approx 720$~keV.

Excitingly, we find that we can probe unexplored Higgsino parameter space even with a much larger $^{238}\text{U}$ background, especially with the HIBM method. For a 1~Gyr sample, this can be achieved even if the uranium background increases by 5 orders of magnitude to $C^{238}=3\times 10^{-7}~\gpg$ (dashed curves in Fig.~\ref{fig: SHM_larger_bg}), and similarly for a 50~Myr sample, for $C^{238}=3\times 10^{-8}~\gpg$ (dashed curves in Fig.~\ref{fig: LMC_larger_bg}). Note that the DM velocity tail is irrelevant for smaller $\delta \lesssim 500$~keV, and Fig.~\ref{fig: SHM_larger_bg} and \ref{fig: LMC_larger_bg} differ only in the age of the mineral in this regime.

Notice that the HIBM method becomes superior to SAXs as backgrounds increase, with the crossover point occurring at about $C^{238}=3\times 10^{-8}~\gpg$. The reason can be seen by comparing Figs.~\ref{fig: obs_spectra} and \ref{fig: obs_spectra_HIBM}. The signal gets exponentially smaller at longer track lengths, whereas the neutron background is mostly flat. When the signal is larger than the background, these features are irrelevant and the SAXs method is preferred due to its larger processing volume ($60~\text{cm}^3$ vs $6~
\text{mm}^3$). As the neutron background gets larger, the lower resolution of the SAXs method, $\sigma_x=15$~nm, smears out the features of the signal into longer track lengths, whereas the HIBM's better resolution of $\sigma_x=1$~nm preserves the features of the signal at short track lengths.

For a 1~Gyr sample, the largest tolerable uranium concentration for the Higgsino is about one part per million (see Fig.~\ref{fig: SHM_larger_bg}), comparable to that of a typical mineral in the Earth's crust~\cite{paleodetectors_details_2018}. We have thus opened up the possibility of using \textit{radio-impure} minerals as paleodetectors.
Furthermore, being able to tolerate such a large uranium background implies that going to 5~km in depth is no longer necessary. Recall that at 5~km depth, the cosmogenic neutron flux becomes negligible compared to radiogenic neutrons at the level of $C^{238}=10^{-10}~\gpg$. Therefore, minerals from the deepest existing laboratories at $\sim 2$~km depth are not viable for conventional paleodetectors, since the cosmogenic neutron flux there is $10^4$ times larger than that at 5 km depth~\cite{paleodetectors_2018}.

Conversely, if we can tolerate $10^4$ times larger radioactivity and still probe new Higgsino parameter space, then existing laboratories at depths of 2~km become viable, opening up vastly more accessible sample sources. Of course, probing larger $\delta$ requires progressively lower backgrounds, but there is a long way to go between 2~km and 5~km over which one can interpolate, and shallower sources are practically more accessible than deeper sources.

Finally, aside from the Higgsino target cross section, note that for both the SHM (Fig.~\ref{fig: SHM_larger_bg}) and the LMC (Fig.~\ref{fig: LMC_larger_bg}), the largest splitting reach is the same as the corresponding curves in Fig.~\ref{fig: projection}. Even samples with uranium concentrations as large as  $C^{238} = 3\times 10^{-7}~\gpg$ can still probe large swaths of unconstrained parameter space.

\begin{figure*}[t]
    \centering
    
    \begin{subfigure}{0.49\textwidth}
        \centering
        \includegraphics[width=\textwidth]{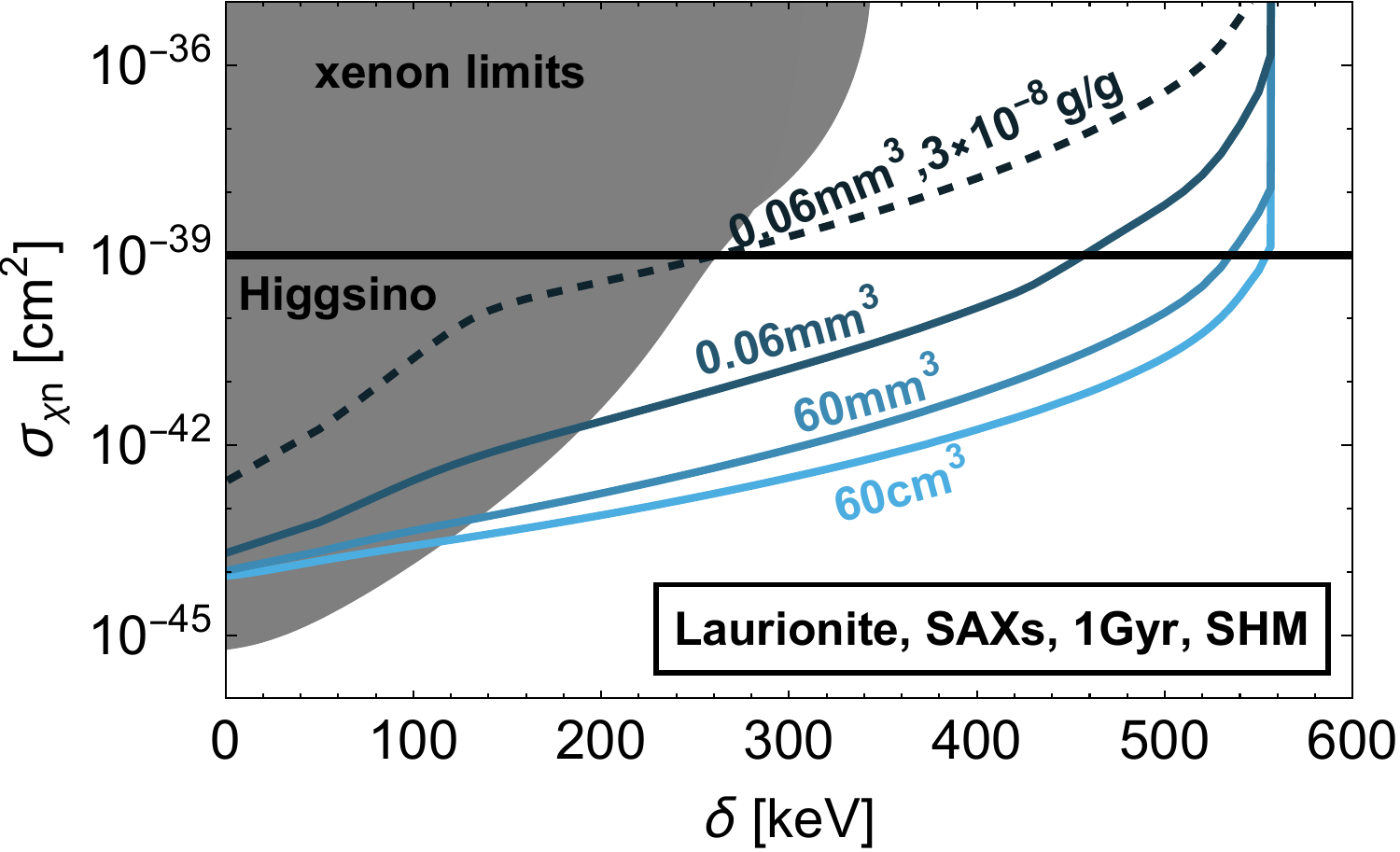}
        \caption{}
        \label{fig: projection_SHM_masses}
    \end{subfigure}
    \hfill
    \begin{subfigure}{0.49\textwidth}
        \centering
        \includegraphics[width=\textwidth]{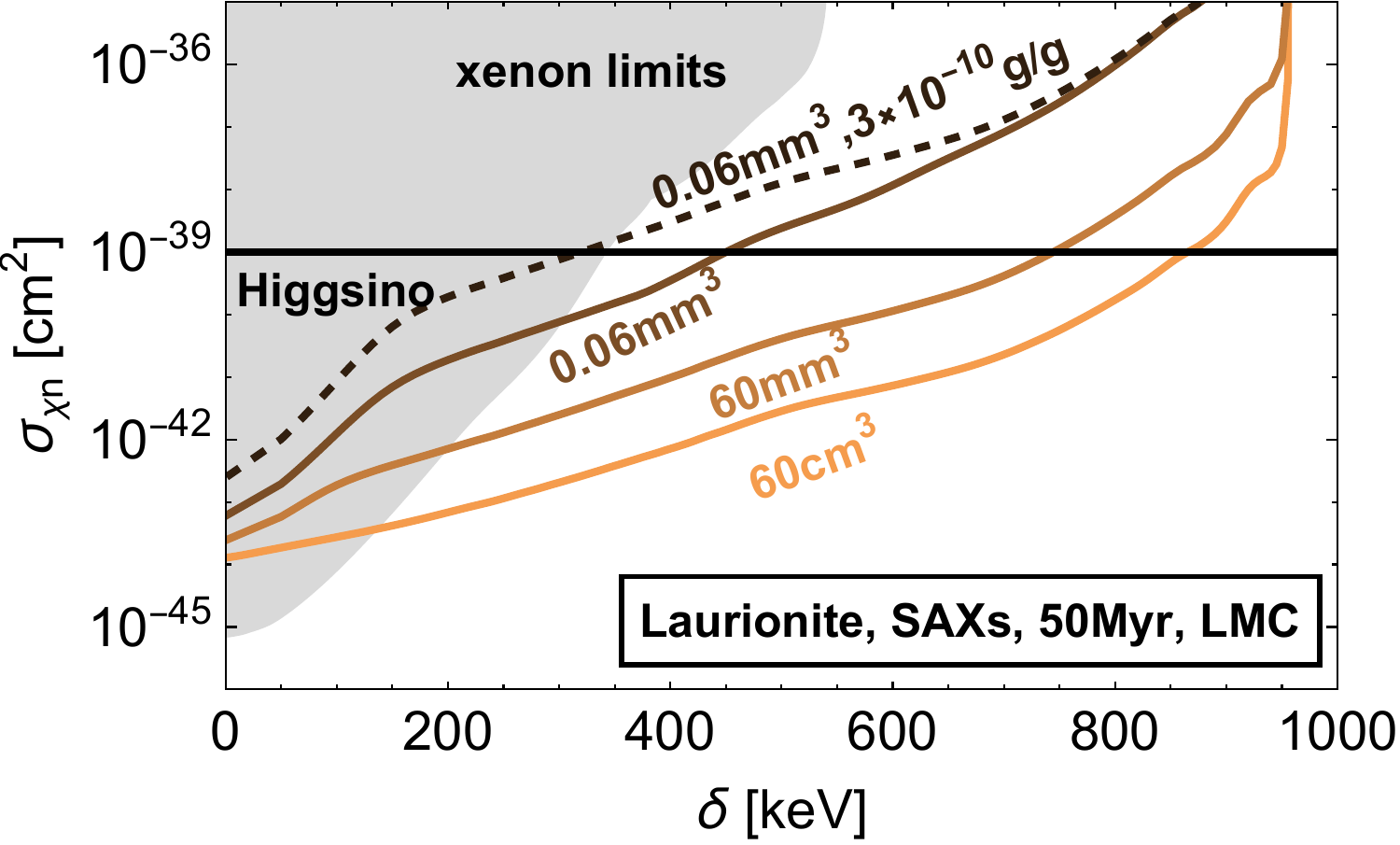}
        \caption{}
        \label{fig: projection_LMC_masses}
    \end{subfigure}
    
    \caption{Same as Fig.~\ref{fig: projection} with resolution $\sigma_x=15$~nm relevant for Small Angle X-ray scattering (SAXs), but we show the effects of progressively smaller sample volumes. Dashed lines show that we can compete with existing Higgsino limits even after reducing the sample volume by a factor of $10^6$, in addition to increasing the uranium background from the benchmark value $C^{238}=3\times 10^{-12}~\gpg$ to (a) $C^{238}=3\times 10^{-8}~\gpg$ for 1~Gyr; (b) $C^{238}=3\times 10^{-10}~\gpg$ for 50~Myr. 
    }
    \label{fig: projection_masses}
\end{figure*}

\subsection{Smaller exposure}
\label{sec: projection_smaller_exposure}

We demonstrate in Fig.~\ref{fig: projection_masses} that the Higgsino signal is so large that we can afford to reduce the target volume by a factor of $10^6$ relative to the benchmark and still probe significantly into unconstrained parameter space. In fact, the dashed curve shows we can take an additional $10^4$ times larger $^{238}\text{U}$ background on top of the million-times smaller target volume\footnote{We show this rather than an even smaller volume because it is getting impractically small.} and still be comparable to existing limits for a 1~Gyr sample; the same is true for $10^2$ times larger $^{238}\text{U}$ background for a 50~Myr sample.

This stands in stark contrast to the elastic WIMP parameter space targeted by conventional paleodetectors, where the unconstrained cross sections are much smaller and the sensitivity to available parameter space hinges on the capacity to read out large samples.

Another practical advantage that comes with the smaller exposure requirement is a proportional relaxation of the data processing. The paleodetector program generically needs to contend with a formidable requirement on data processing, with the SAXs benchmark considered here ($60~\text{cm}^3$ at $\sigma_x=15~\text{nm}$ resolution) requiring about $\sim 10^{20}$~Bytes, approaching high-luminosity
LHC level~\cite{Hedges:2026pgf}. With our approach, one could start with a million-fold smaller data volumes while still probing new Higgsino parameter space.

\section{Conclusion}
\label{sec: conclusion}

We have proposed heavy-element paleodetectors capable of pushing far deeper into inelastic dark matter parameter space than previously possible. Because the kinematic threshold set by the mass splitting $\delta$ scales linearly with the target nucleus mass, paleodetectors containing heavy elements such as Pb probe large $\delta$. A prime example is the relic-mass Higgsino, whose Z-mediated cross section leaves $\delta$ as the only unknown parameter. In the most optimistic scenario, heavy-element paleodetectors could probe Higgsino mass splittings up to $\delta \simeq 920$~keV, thus covering a significant region of the remaining parameter space for this classic supersymmetric WIMP dark matter.

We have identified brine precipitates as a possible geological source of radiopure minerals containing heavy elements, distinct from sources considered in existing paleodetector literature. We have found that paleodetectors are uniquely sensitive to ancient fast DM, which may have contained a faster population than the present-day distribution due to the impact of the LMC 50~Myr ago. This history also favors somewhat younger minerals, contrary to the usual expectation. 

We have found that, even before finding the optimal minerals to target deep into Higgsino parameter space, it is already possible to discover the Higgsino using accessible, suboptimal samples with radioactivity typical of the Earth's crust and from depths of 2~km, rather than radiopure samples at 5~km depths, thus alleviating the difficulty of finding suitable samples. This exciting possibility is unique to inelastic dark matter, and differs substantially from the experimental scenarios considered thus far in the paleodetector literature. Furthermore, it is possible to probe new parameter space even with much smaller sample volumes, thus alleviating the formidable requirements on data processing.

This work also opens up a new frontier for inelastic DM: the study of the history of the DM velocity tail, beyond the impact of the LMC, perhaps by simulating earlier MW mergers. Such simulations could potentially find even faster ancient DM, informing the preferred age of mineral targets, and pushing the sensitivity to larger mass splittings. We encourage the astrophysics community to study this, especially since many simulations already start with early initial conditions and time-evolve to the present day, but publish only the present-day distribution.

This work is complementary to our proposal in Ref.~\cite{EnhancingHiggsinoDM_2024} to place a large volume of heavy elements near a neutrino detector and detect the subsequent decay of the excited state via a photon. In comparison to this luminous detection method, paleodetectors have different technological challenges, but are sensitive to a possibly faster ancient DM velocity distribution, which is inaccessible to real-time detectors. 

A natural extension of this work is to consider the sensitivity of heavy-element paleodetectors to electroweak WIMPs in other representations of the standard model gauge group, as well as WIMP masses that differ from the freeze-out mass due to non-standard cosmologies. Interestingly, while fermionic electroweak WIMPs can be probed by either luminous detection or paleodetectors, scalar electroweak WIMPs have parametrically longer decay lengths than fermionic ones~\cite{Bottaro:2022one}, making them inaccessible to luminous detection~\cite{luminous_complex_WIMP}. Hence, heavy-element paleodetectors are the only known direct detection method that could probe scalar electroweak WIMPs in this regime.

\acknowledgments
We thank Masha Baryakhtar, Emilie LaVoie-Ingram, Patrick Stengel, and Xiuyuan Zhang for helpful discussions. S.~W. is supported in part by the U.S. Department of Energy Office of Science under Award Number DE-SC0024375, and by the Gordon and Betty Moore Foundation through Grant GBMF13898 to the University of 
Washington.
This work was supported in part by NSF Grant No. PHY-2310429, NSF Grant No. PHY-2515007, Simons Investigator Award No. 824870, The University of Delaware Research Foundation and the John Templeton Foundation Award No. 63595.

\appendix

\section{Stopping power}
\label{app: stopping power}
For recoil energies less than 1 MeV/nucleon, the stopping power can be written as the sum of elastic nuclear interactions and electronic excitations~\cite{dEdx_Averback_1998},
\begin{align}
    \frac{dE}{dx} = \left( \frac{dE}{dx} \right)_n + \left( \frac{dE}{dx} \right)_e ~.
\end{align}
We use the following semi-analytic approximations~\cite{dEdx_Biersack_1977}.

For an incident atom with mass $M_1$ and lab-frame energy $E$ interacting with a target atom of mass $M_2$, define the reduced energy $\epsilon$ as
\begin{align}
    \epsilon \equiv \frac{a E}{Z_1 Z_2 \alpha} \frac{M_2}{M_1+M_2} ~,
\end{align}
where $\alpha =1/137$ is the fine structure constant, and the screening length $a$ is defined as~\cite{dEdx_Lindhard_1968}
\begin{align}
    a \equiv \frac{0.8853~a_0}{(Z_1^{2/3}+Z_2^{2/3})^{1/2}} ~,
\end{align}
with $a_0 \approx 0.5 \AA$ being the Bohr radius.
Then it can be shown that~\cite{dEdx_Biersack_1977} 
\begin{align}
\frac{dE}{dx} = \frac{\pi a^2 \gamma n}{C} S(\epsilon) ~,
\end{align}
where $n$ is the target atomic number density, $C \equiv \epsilon/E$, $\gamma \equiv 4 M_1 M_2/(M_1+M_2)^2$, and $S(\epsilon)=S_n(\epsilon)+S_e(\epsilon)$, with
\begin{align}
    S_n(\epsilon) &= \frac{0.5 \log(1+\epsilon)}{\epsilon+A \epsilon^B} \\
    S_e(\epsilon) &= k \sqrt{\epsilon}~,
\end{align}
with the fitted constants $A=0.14120$, $B=0.42059$, and $k = 0.15$. For a composite material, we sum over the different constituent elements $A$~\cite{paleodetectors_details_2018},
\begin{align}
    \frac{dE}{dx} = \sum_A \left. g_A \frac{dE}{dx} \right|_A ~,
\end{align}
where $g_A$ is the number of atoms $A$ appearing in the chemical formula of the material. We verified that this approximation is in good agreement with the numerical result found in Ref.~\cite{paleodetectors_2018}.

\section{Other heavy-element minerals}
\label{sec:other_minerals}
Other Pb-bearing minerals that could form in a chlorine-rich environment include Cottunite $(\text{PbCl}_2)$, Penfieldite $(\text{Pb}_2\text{Cl}_3\text{OH})$, Fiedlerite $[\text{Pb}_3\text{Cl}_4(\text{OH})_2]$, Paralaurionite\footnote{Dimorph of Laurionite.} $(\text{PbOHCl})$, Mendipite $(\text{Pb}_3\text{O}_2\text{Cl}_2)$, Blixite $[\text{Pb}_2\text{Cl}(\text{OH})_3]$, and Litharge $(\text{PbO})$~\cite{Edwards_Gillard_Williams_Pollard_1992}.

Laurionite is an example of a class of minerals known as halides, whose chemical formulas contain a halogen and a metal. They are generally soluble in water~\cite{Sorrel1973Halides}, and thus could be compatible with BP formation. In the Nickel-Strunz Classification (Version 10)~\cite{webmineral_nickel_strunz,strunz1941mineralogische}, the categories 03.DC and 03.DB contain many halides with Pb and Bi in their chemical formulas. We hypothesize that these categories contain other promising candidates that can form from BPs, but leave further investigation to future work.

\section{Potential mineral locations}
\label{sec: mineral_locations}
In the 1980s, the U.S. Department of Energy drilled deep wells into about a dozen aquifers in the Northern Gulf of Mexico in an effort to evaluate their potential for producing energy, and Kraemer~et~al. studied the extracted samples of geothermal brines~\cite{Kraemer_lowU_1981,Kraemer_lowU_1986,geopressure_review_1992}. Among the dozen samples tested, most have uranium concentrations within an order of magnitude of the value $C^{238}_{\rm BP} = 3\times 10^{-12}~\gpg$. The four aquifers that reach this lowest reported concentration are Pleasant Bayou, Texas; L. R. Sweezy, Louisiana; Amoco Fee, Louisiana; and Gladys McCall, Louisiana.

In fact, we find that these U.S. Gulf Coast aquifers not only serve as existence proofs of radiopure brines, but could contain potential heavy-element paleodetector targets, since they satisfy other relevant criteria. First, they have depths between 4 and 5~km~\cite{Kraemer_lowU_1981}, satisfying the requirement of shielding from cosmic rays. Second, some of these aquifers are Oligocene (23--34 Myr ago) in age~\cite{Kraemer_lowU_1981}, within an $\mathcal{O}(1)$ factor of the $\sim 50$~Myr favored by the LMC scenario. The earliest mineral precipitates formed in these aquifers thus have the desired age. Third, some of these aquifers are reported to have high concentrations of Pb and Cl~\cite{Kraemer_lowU_1986}, e.g., Pleasant Bayou has $C^{\rm Pb}=1.1 \times 10^{-6}~\gpg$, $C^{\rm Cl}=8\times 10^{-2}~\gpg$, and pH = 6.5. It could well be that ideal samples of Laurionite are present in these wells.

\bibliography{refs}

@article{Chang:2010en,
    author = "Chang, Spencer and Weiner, Neal and Yavin, Itay",
    title = "{Magnetic Inelastic Dark Matter}",
    eprint = "1007.4200",
    archivePrefix = "arXiv",
    primaryClass = "hep-ph",
    doi = "10.1103/PhysRevD.82.125011",
    journal = "Phys. Rev. D",
    volume = "82",
    pages = "125011",
    year = "2010"
}

@article{Bramante:2016rdh,
    author = "Bramante, Joseph and Fox, Patrick J. and Kribs, Graham D. and Martin, Adam",
    title = "{Inelastic frontier: Discovering dark matter at high recoil energy}",
    eprint = "1608.02662",
    archivePrefix = "arXiv",
    primaryClass = "hep-ph",
    reportNumber = "FERMILAB-PUB-16-301-T",
    doi = "10.1103/PhysRevD.94.115026",
    journal = "Phys. Rev. D",
    volume = "94",
    number = "11",
    pages = "115026",
    year = "2016"
}

@article{Baryakhtar:2020rwy,
    author = "Baryakhtar, Masha and Berlin, Asher and Liu, Hongwan and Weiner, Neal",
    title = "{Electromagnetic signals of inelastic dark matter scattering}",
    eprint = "2006.13918",
    archivePrefix = "arXiv",
    primaryClass = "hep-ph",
    doi = "10.1007/JHEP06(2022)047",
    journal = "JHEP",
    volume = "06",
    pages = "047",
    year = "2022"
}

@article{Tucker-Smith:2001myb,
    author = "Tucker-Smith, David and Weiner, Neal",
    title = "{Inelastic dark matter}",
    eprint = "hep-ph/0101138",
    archivePrefix = "arXiv",
    reportNumber = "UCB-PTH-00-43, LBNL-47234, UW-PT-00-17",
    doi = "10.1103/PhysRevD.64.043502",
    journal = "Phys. Rev. D",
    volume = "64",
    pages = "043502",
    year = "2001"
}

@article{Barn,
    author = "Digman, Matthew C. and Cappiello, Christopher V. and Beacom, John F. and Hirata, Christopher M. and Peter, Annika H. G.",
    title = "{Not as big as a barn: Upper bounds on dark matter-nucleus cross sections}",
    eprint = "1907.10618",
    archivePrefix = "arXiv",
    primaryClass = "hep-ph",
    doi = "10.1103/PhysRevD.100.063013",
    journal = "Phys. Rev. D",
    volume = "100",
    number = "6",
    pages = "063013",
    year = "2019",
    note = "[Erratum: Phys.Rev.D 106, 089902 (2022)]"
}

@article{Feldstein:2010su,
    author = "Feldstein, Brian and Graham, Peter W. and Rajendran, Surjeet",
    title = "{Luminous dark matter}",
    eprint = "1008.1988",
    archivePrefix = "arXiv",
    primaryClass = "hep-ph",
    reportNumber = "MIT-CTP-4172",
    doi = "10.1103/PhysRevD.82.075019",
    journal = "Phys. Rev. D",
    volume = "82",
    pages = "075019",
    year = "2010"
}

@article{LMC,
    author = "Smith-Orlik, Adam and others",
    title = "{The impact of the Large Magellanic Cloud on dark matter direct detection signals}",
    eprint = "2302.04281",
    archivePrefix = "arXiv",
    primaryClass = "astro-ph.GA",
    doi = "10.1088/1475-7516/2023/10/070",
    journal = "JCAP",
    volume = "10",
    pages = "070",
    year = "2023"
}

@article{Luminous,
    author = "Eby, Joshua and Fox, Patrick J. and Harnik, Roni and Kribs, Graham D.",
    title = "{Luminous signals of inelastic dark matter in large detectors}",
    eprint = "1904.09994",
    archivePrefix = "arXiv",
    primaryClass = "hep-ph",
    reportNumber = "FERMILAB-PUB-19-147-T",
    doi = "10.1007/JHEP09(2019)115",
    journal = "JHEP",
    volume = "09",
    pages = "115",
    year = "2019"
}

@article{MagneticInelasticDM,
    author = "Eby, Joshua and Fox, Patrick J. and Kribs, Graham D.",
    title = "{Earth-catalyzed detection of magnetic inelastic dark matter with photons in large underground detectors}",
    eprint = "2312.08478",
    archivePrefix = "arXiv",
    primaryClass = "hep-ph",
    reportNumber = "FERMILAB-PUB-23-781-T, IPMU23-0046",
    doi = "10.1007/JHEP06(2024)165",
    journal = "JHEP",
    volume = "06",
    pages = "165",
    year = "2024"
}

@article{LMCHighSpeed,
    author = "Besla, Gurtina and Peter, Annika and Garavito-Camargo, Nicolas",
    title = "{The highest-speed local dark matter particles come from the Large Magellanic Cloud}",
    eprint = "1909.04140",
    archivePrefix = "arXiv",
    primaryClass = "astro-ph.GA",
    doi = "10.1088/1475-7516/2019/11/013",
    journal = "JCAP",
    volume = "11",
    pages = "013",
    year = "2019"
}

@article{Original_SHM,
  title = {Detecting cold dark-matter candidates},
  author = {Drukier, Andrzej K. and Freese, Katherine and Spergel, David N.},
  journal = {Phys. Rev. D},
  volume = {33},
  issue = {12},
  pages = {3495--3508},
  numpages = {0},
  year = {1986},
  month = {Jun},
  publisher = {American Physical Society},
  doi = {10.1103/PhysRevD.33.3495},
  url = {https://link.aps.org/doi/10.1103/PhysRevD.33.3495}
}

@article{SMH_Deviation_2,
    author = "Vogelsberger, Mark and Helmi, A. and Springel, Volker and White, Simon D. M. and Wang, Jie and Frenk, Carlos S. and Jenkins, Adrian and Ludlow, A. D. and Navarro, Julio F.",
    title = "{Phase-space structure in the local dark matter distribution and its signature in direct detection experiments}",
    eprint = "0812.0362",
    archivePrefix = "arXiv",
    primaryClass = "astro-ph",
    doi = "10.1111/j.1365-2966.2009.14630.x",
    journal = "Mon. Not. Roy. Astron. Soc.",
    volume = "395",
    pages = "797--811",
    year = "2009"
}

@article{SMH_Deviation_5,
    author = "Mao, Yao-Yuan and Strigari, Louis E. and Wechsler, Risa H.",
    title = "{Connecting Direct Dark Matter Detection Experiments to Cosmologically Motivated Halo Models}",
    eprint = "1304.6401",
    archivePrefix = "arXiv",
    primaryClass = "astro-ph.CO",
    doi = "10.1103/PhysRevD.89.063513",
    journal = "Phys. Rev. D",
    volume = "89",
    number = "6",
    pages = "063513",
    year = "2014"
}

@article{LMC_Similar,
   title={Effects on the local dark matter distribution due to the large magellanic cloud},
   volume={513},
   ISSN={1745-3933},
   url={http://dx.doi.org/10.1093/mnrasl/slac031},
   DOI={10.1093/mnrasl/slac031},
   number={1},
   journal={Monthly Notices of the Royal Astronomical Society: Letters},
   publisher={Oxford University Press (OUP)},
   author={Donaldson, Katelin and Petersen, Michael S and Peñarrubia, Jorge},
   year={2022},
   month=mar, pages={46–51} }

@article{Auriga,
    author = "Grand, Robert J. J. and G\'omez, Facundo A. and Marinacci, Federico and Pakmor, Ruediger and Springel, Volker and Campbell, David J. R. and Frenk, Carlos S. and Jenkins, Adrian and White, Simon D. M.",
    title = "{The Auriga Project: the properties and formation mechanisms of disc galaxies across cosmic time}",
    eprint = "1610.01159",
    archivePrefix = "arXiv",
    primaryClass = "astro-ph.GA",
    doi = "10.1093/mnras/stx071",
    journal = "Mon. Not. Roy. Astron. Soc.",
    volume = "467",
    number = "1",
    pages = "179--207",
    year = "2017"
}

@misc{JUNO_Conceptual,
    author = "Djurcic, Zelimir and others",
    collaboration = "JUNO",
    title = "{JUNO Conceptual Design Report}",
    eprint = "1508.07166",
    archivePrefix = "arXiv",
    primaryClass = "physics.ins-det",
    month = "8",
    year = "2015"
}

@article{acme2018improved,
	title = {Improved limit on the electric dipole moment of the electron},
	volume = {562},
	issn = {1476-4687},
	url = {https://doi.org/10.1038/s41586-018-0599-8},
	doi = {10.1038/s41586-018-0599-8},
	number = {7727},
	journal = {Nature},
	author = {Andreev, V. and Ang, D. G. and DeMille, D. and Doyle, J. M. and Gabrielse, G. and Haefner, J. and Hutzler, N. R. and Lasner, Z. and Meisenhelder, C. and O’Leary, B. R. and Panda, C. D. and West, A. D. and West, E. P. and Wu {(ACME Collaboration)}, X.},
	month = oct,
	year = {2018},
	pages = {355--360},
}

@article{PandaX4T2021,
  title = {Dark Matter Search Results from the PandaX-4T Commissioning Run},
  author = {Meng \textit{et al.}, Yue},
  collaboration = {PandaX-4T Collaboration},
  journal = {Phys. Rev. Lett.},
  volume = {127},
  issue = {26},
  pages = {261802},
  numpages = {8},
  year = {2021},
  month = {Dec},
  publisher = {American Physical Society},
  doi = {10.1103/PhysRevLett.127.261802},
  url = {https://link.aps.org/doi/10.1103/PhysRevLett.127.261802}
}

@article{Co:2021ion,
    author = "Co, Raymond T. and Sheff, Benjamin and Wells, James D.",
    title = "{Race to find split Higgsino dark matter}",
    eprint = "2105.12142",
    archivePrefix = "arXiv",
    primaryClass = "hep-ph",
    reportNumber = "LCTP-21-11, UMN-TH-4016/21, FTPI-MINN-21/09",
    doi = "10.1103/PhysRevD.105.035012",
    journal = "Phys. Rev. D",
    volume = "105",
    number = "3",
    pages = "035012",
    year = "2022"
}

@article{LastWIMP,
    author = "Krall, Rebecca and Reece, Matthew",
    title = "{Last Electroweak WIMP Standing: Pseudo-Dirac Higgsino Status and Compact Stars as Future Probes}",
    eprint = "1705.04843",
    archivePrefix = "arXiv",
    primaryClass = "hep-ph",
    doi = "10.1088/1674-1137/42/4/043105",
    journal = "Chin. Phys. C",
    volume = "42",
    number = "4",
    pages = "043105",
    year = "2018"
}

@article{Rodd:2024qsi,
    author = "Rodd, Nicholas L. and Safdi, Benjamin R. and Xu, Weishuang Linda",
    title = "{CTA and SWGO can discover Higgsino dark matter annihilation}",
    eprint = "2405.13104",
    archivePrefix = "arXiv",
    primaryClass = "hep-ph",
    doi = "10.1103/PhysRevD.110.043003",
    journal = "Phys. Rev. D",
    volume = "110",
    number = "4",
    pages = "043003",
    year = "2024"
}

@article{CTAConsortium:2010umy,
    author = "Actis, M. and others",
    collaboration = "CTA Consortium",
    title = "{Design concepts for the Cherenkov Telescope Array CTA: An advanced facility for ground-based high-energy gamma-ray astronomy}",
    eprint = "1008.3703",
    archivePrefix = "arXiv",
    primaryClass = "astro-ph.IM",
    doi = "10.1007/s10686-011-9247-0",
    journal = "Exper. Astron.",
    volume = "32",
    pages = "193--316",
    year = "2011"
}

@article{Rinchiuso:2020skh,
    author = "Rinchiuso, Lucia and Macias, Oscar and Moulin, Emmanuel and Rodd, Nicholas L. and Slatyer, Tracy R.",
    title = "{Prospects for detecting heavy WIMP dark matter with the Cherenkov Telescope Array: The Wino and Higgsino}",
    eprint = "2008.00692",
    archivePrefix = "arXiv",
    primaryClass = "astro-ph.HE",
    reportNumber = "MIT-CTP 5120, IRFU-20-13",
    doi = "10.1103/PhysRevD.103.023011",
    journal = "Phys. Rev. D",
    volume = "103",
    number = "2",
    pages = "023011",
    year = "2021"
}

@article{aaboud2018search,
  title={Search for electroweak production of supersymmetric states in scenarios with compressed mass spectra at s= 13 TeV with the ATLAS detector},
  author={Aaboud, Morad and Aad, Georges and Abbott, Brad and Abdinov, Ovsat and Abeloos, Baptiste and Abidi, Syed Haider and AbouZeid, OS and Abraham, Nadine L and Abramowicz, Halina and Abreu, Henso and others},
  journal={Physical Review D},
  volume={97},
  number={5},
  pages={052010},
  year={2018},
  publisher={APS}
}

@article{inelastic_frontier_Bramante_2016,
    author = "Bramante, Joseph and Fox, Patrick J. and Kribs, Graham D. and Martin, Adam",
    title = "{Inelastic frontier: Discovering dark matter at high recoil energy}",
    eprint = "1608.02662",
    archivePrefix = "arXiv",
    primaryClass = "hep-ph",
    reportNumber = "FERMILAB-PUB-16-301-T",
    doi = "10.1103/PhysRevD.94.115026",
    journal = "Phys. Rev. D",
    volume = "94",
    number = "11",
    pages = "115026",
    year = "2016"
}

@article{XENON1T,
    author = "Aprile, E. and others",
    collaboration = "XENON",
    title = "{Dark Matter Search Results from a One Ton-Year Exposure of XENON1T}",
    eprint = "1805.12562",
    archivePrefix = "arXiv",
    primaryClass = "astro-ph.CO",
    doi = "10.1103/PhysRevLett.121.111302",
    journal = "Phys. Rev. Lett.",
    volume = "121",
    number = "11",
    pages = "111302",
    year = "2018"
}

@article{EnhancingHiggsinoDM_2024,
    author = "Graham, Peter W. and Ramani, Harikrishnan and Wong, Samuel S. Y.",
    title = "{Enhancing direct detection of Higgsino dark matter}",
    eprint = "2409.07768",
    archivePrefix = "arXiv",
    primaryClass = "hep-ph",
    doi = "10.1103/PhysRevD.111.055030",
    journal = "Phys. Rev. D",
    volume = "111",
    number = "5",
    pages = "055030",
    year = "2025"
}

@article{dEdx_Biersack_1977,
  title = {Calculations of nuclear stopping, ranges, and straggling in the low-energy region},
  author = {Wilson, W. D. and Haggmark, L. G. and Biersack, J. P.},
  journal = {Phys. Rev. B},
  volume = {15},
  issue = {5},
  pages = {2458--2468},
  numpages = {0},
  year = {1977},
  month = {Mar},
  publisher = {American Physical Society},
  doi = {10.1103/PhysRevB.15.2458},
  url = {https://link.aps.org/doi/10.1103/PhysRevB.15.2458}
}

@incollection{dEdx_Averback_1998,
title = {Displacement Damage in Irradiated Metals and Semiconductors},
editor = {Henry Ehrenreich and Frans Spaepen},
series = {Solid State Physics},
publisher = {Academic Press},
volume = {51},
pages = {281-402},
year = {1998},
booktitle = {Solid State Physics},
issn = {0081-1947},
doi = {https://doi.org/10.1016/S0081-1947(08)60193-9},
url = {https://www.sciencedirect.com/science/article/pii/S0081194708601939},
author = {R.S. AVERBACK and T. DIAZ {DE LA RUBIA}}
}

@article{dEdx_Lindhard_1968,
  author       = {Lindhard, J and Nielsen, V and Scharff, M},
  title        = {APPROXIMATION METHOD IN CLASSICAL SCATTERING BY SCREENED COULOMB FIELDS.},
  url          = {https://www.osti.gov/biblio/4503382},
  journal      = {Kgl. Dan. Vidensk. Selsk., Mat.-Fys. Medd., 36:  31p(1968).},
  place        = {Denmark},
  year         = {1967},
  month        = {12}}

@article{paleodetectors_2018,
    author = "Baum, Sebastian and Drukier, Andrzej K. and Freese, Katherine and G{\'o}rski, Maciej and Stengel, Patrick",
    title = "{Searching for Dark Matter with Paleo-Detectors}",
    eprint = "1806.05991",
    archivePrefix = "arXiv",
    primaryClass = "astro-ph.CO",
    reportNumber = "NORDITA-2018-043, LCTP-18-15",
    doi = "10.1016/j.physletb.2020.135325",
    journal = "Phys. Lett. B",
    volume = "803",
    pages = "135325",
    year = "2020"
}

@article{SAX_resolution_2014,
	title = {X-ray ptychographic computed tomography at 16 nm isotropic {3D} resolution},
	volume = {4},
	issn = {2045-2322},
	url = {https://doi.org/10.1038/srep03857},
	doi = {10.1038/srep03857},
	number = {1},
	journal = {Scientific Reports},
	author = {Holler, M. and Diaz, A. and Guizar-Sicairos, M. and Karvinen, P. and Färm, Elina and Härkönen, Emma and Ritala, Mikko and Menzel, A. and Raabe, J. and Bunk, O.},
	month = jan,
	year = {2014},
	pages = {3857},
}

@article{Math_of_DM_Lewin_Smith_1995,
    author = "Lewin, J. D. and Smith, P. F.",
    title = "{Review of mathematics, numerical factors, and corrections for dark matter experiments based on elastic nuclear recoil}",
    reportNumber = "RAL-TR-95-024",
    doi = "10.1016/S0927-6505(96)00047-3",
    journal = "Astropart. Phys.",
    volume = "6",
    pages = "87--112",
    year = "1996"
}

@article{magic_number_Mayer_1948,
  title = {On Closed Shells in Nuclei},
  author = {Mayer, Maria G.},
  journal = {Phys. Rev.},
  volume = {74},
  issue = {3},
  pages = {235--239},
  numpages = {0},
  year = {1948},
  month = {Aug},
  publisher = {American Physical Society},
  doi = {10.1103/PhysRev.74.235},
  url = {https://link.aps.org/doi/10.1103/PhysRev.74.235}
}

@article{LMC_pericenter_time,
   title={The Orbital Histories of Magellanic Satellites Using Gaia DR2 Proper Motions},
   volume={893},
   ISSN={1538-4357},
   url={http://dx.doi.org/10.3847/1538-4357/ab7b75},
   DOI={10.3847/1538-4357/ab7b75},
   number={2},
   journal={The Astrophysical Journal},
   publisher={American Astronomical Society},
   author={Patel, Ekta and Kallivayalil, Nitya and Garavito-Camargo, Nicolas and Besla, Gurtina and Weisz, Daniel R. and van der Marel, Roeland P. and Boylan-Kolchin, Michael and Pawlowski, Marcel S. and Gómez, Facundo A.},
   year={2020},
   month=apr, pages={121} }

@article{paleodetectors_details_2018,
    author = "Drukier, Andrzej K. and Baum, Sebastian and Freese, Katherine and G{\'o}rski, Maciej and Stengel, Patrick",
    title = "{Paleo-detectors: Searching for Dark Matter with Ancient Minerals}",
    eprint = "1811.06844",
    archivePrefix = "arXiv",
    primaryClass = "astro-ph.CO",
    reportNumber = "NORDITA-2018-117, LCTP-18-25",
    doi = "10.1103/PhysRevD.99.043014",
    journal = "Phys. Rev. D",
    volume = "99",
    number = "4",
    pages = "043014",
    year = "2019"
}

@article{paleodetectors_new_2021,
    author = "Baum, Sebastian and Edwards, Thomas D. P. and Freese, Katherine and Stengel, Patrick",
    title = "{New Projections for Dark Matter Searches with Paleo-Detectors}",
    eprint = "2106.06559",
    archivePrefix = "arXiv",
    primaryClass = "astro-ph.CO",
    doi = "10.3390/instruments5020021",
    journal = "Instruments",
    volume = "5",
    number = "2",
    pages = "21",
    year = "2021"
}

@article{paleodetectors_spectral_2018,
    author = "Edwards, Thomas D. P. and Kavanagh, Bradley J. and Weniger, Christoph and Baum, Sebastian and Drukier, Andrzej K. and Freese, Katherine and G{\'o}rski, Maciej and Stengel, Patrick",
    title = "{Digging for dark matter: Spectral analysis and discovery potential of paleo-detectors}",
    eprint = "1811.10549",
    archivePrefix = "arXiv",
    primaryClass = "hep-ph",
    reportNumber = "NORDITA-2018-119; LCTP-18-26",
    doi = "10.1103/PhysRevD.99.043541",
    journal = "Phys. Rev. D",
    volume = "99",
    number = "4",
    pages = "043541",
    year = "2019"
}

@article{paleodetectors_refine_2025,
    author = "Fung, Audrey and Lucas, Thalles and Balogh, Levente and Leybourne, Matthew and Vincent, Aaron C.",
    title = "{Refining the sensitivity of new physics searches with ancient minerals}",
    eprint = "2504.08885",
    archivePrefix = "arXiv",
    primaryClass = "hep-ph",
    doi = "10.1103/clvr-mhx5",
    journal = "Phys. Rev. D",
    volume = "112",
    number = "4",
    pages = "043040",
    year = "2025"
}

@article{Mica_Snowden_1995,
    author = "Snowden-Ifft, D. P. and Freeman, E. S. and Price, P. B.",
    title = "{Limits on dark matter using ancient mica}",
    doi = "10.1103/PhysRevLett.74.4133",
    journal = "Phys. Rev. Lett.",
    volume = "74",
    pages = "4133--4136",
    year = "1995"
}

@article{mineral_resistivity_1950,
	title = {Thermoelectric power and electrical resistivity of minerals*},
	volume = {35},
	issn = {0003-004X},
	number = {7-8},
	journal = {American Mineralogist},
	author = {Telkes, Maria},
	month = aug,
	year = {1950},
	note = {\_eprint: https://pubs.geoscienceworld.org/msa/ammin/article-pdf/35/7-8/536/4244479/am-1950-536.pdf},
	pages = {536--555},
}

@article{Nagata:2014wma,
    author = "Nagata, Natsumi and Shirai, Satoshi",
    title = "{Higgsino Dark Matter in High-Scale Supersymmetry}",
    eprint = "1410.4549",
    archivePrefix = "arXiv",
    primaryClass = "hep-ph",
    reportNumber = "DESY-14-180, FTPI-MINN-14-37, IPMU14-0320",
    doi = "10.1007/JHEP01(2015)029",
    journal = "JHEP",
    volume = "01",
    pages = "029",
    year = "2015"
}

@article{Hisano:2011cs,
    author = "Hisano, Junji and Ishiwata, Koji and Nagata, Natsumi and Takesako, Tomohiro",
    title = "{Direct Detection of Electroweak-Interacting Dark Matter}",
    eprint = "1104.0228",
    archivePrefix = "arXiv",
    primaryClass = "hep-ph",
    reportNumber = "IPMU-11-0046, ICRR-REPORT-583-2010-16, CALT-68-2824",
    doi = "10.1007/JHEP07(2011)005",
    journal = "JHEP",
    volume = "07",
    pages = "005",
    year = "2011"
}

@article{Hisano:2012wm,
    author = "Hisano, Junji and Ishiwata, Koji and Nagata, Natsumi",
    title = "{Direct Search of Dark Matter in High-Scale Supersymmetry}",
    eprint = "1210.5985",
    archivePrefix = "arXiv",
    primaryClass = "hep-ph",
    reportNumber = "IPMU-12-0187, CALT-68-2891",
    doi = "10.1103/PhysRevD.87.035020",
    journal = "Phys. Rev. D",
    volume = "87",
    pages = "035020",
    year = "2013"
}

@article{Abe:2025lci,
    author = "Abe, Shotaro and Inada, Tomohiro and Moulin, Emmanuel and Rodd, Nicholas L. and Safdi, Benjamin R. and Xu, Weishuang Linda",
    title = "{Discovering the Higgsino at CTAO-North within the Decade}",
    eprint = "2506.08084",
    archivePrefix = "arXiv",
    primaryClass = "hep-ph",
    month = "6",
    journal = {},
    year = "2025"
}

@article{Low:2014cba,
    author = "Low, Matthew and Wang, Lian-Tao",
    title = "{Neutralino dark matter at 14 TeV and 100 TeV}",
    eprint = "1404.0682",
    archivePrefix = "arXiv",
    primaryClass = "hep-ph",
    reportNumber = "EFI-14-6",
    doi = "10.1007/JHEP08(2014)161",
    journal = "JHEP",
    volume = "08",
    pages = "161",
    year = "2014"
}

@article{Kowalska:2018toh,
    author = "Kowalska, Kamila and Sessolo, Enrico Maria",
    title = "{The discreet charm of higgsino dark matter - a pocket review}",
    eprint = "1802.04097",
    archivePrefix = "arXiv",
    primaryClass = "hep-ph",
    doi = "10.1155/2018/6828560",
    journal = "Adv. High Energy Phys.",
    volume = "2018",
    pages = "6828560",
    year = "2018"
}

@article{Bhattiprolu:2025zwt,
    author = "Bhattiprolu, Prudhvi N. and Martin, Stephen P. and Wells, James D.",
    title = "{Chasing higgsino dark matter at colliders in the neutrino fog era}",
    eprint = "2512.12457",
    archivePrefix = "arXiv",
    primaryClass = "hep-ph",
    month = "12",
    year = "2025",
    journal = {},
}

@article{Bottaro:2022one,
    author = "Bottaro, Salvatore and Buttazzo, Dario and Costa, Marco and Franceschini, Roberto and Panci, Paolo and Redigolo, Diego and Vittorio, Ludovico",
    title = "{The last complex WIMPs standing}",
    eprint = "2205.04486",
    archivePrefix = "arXiv",
    primaryClass = "hep-ph",
    reportNumber = "CERN-TH-2022-080",
    doi = "10.1140/epjc/s10052-022-10918-5",
    journal = "Eur. Phys. J. C",
    volume = "82",
    number = "11",
    pages = "992",
    year = "2022"
}

@article
{LZ:2019sgr,
    author = "Akerib, D. S. and others",
    collaboration = "LZ",
    title = "{The LUX-ZEPLIN (LZ) Experiment}",
    eprint = "1910.09124",
    archivePrefix = "arXiv",
    primaryClass = "physics.ins-det",
    reportNumber = "FERMILAB-PUB-19-555-AE-E",
    doi = "10.1016/j.nima.2019.163047",
    journal = "Nucl. Instrum. Meth. A",
    volume = "953",
    pages = "163047",
    year = "2020"
}

@article{Theodosopoulos:2026ehn,
    author = "Theodosopoulos, Dionysios P. and Freese, Katherine and Kelso, Chris and Stengel, Patrick",
    title = "{Projected Sensitivity of Paleo-Detectors to Dark Matter Effective Interactions with Nuclei}",
    eprint = "2603.13629",
    archivePrefix = "arXiv",
    primaryClass = "astro-ph.CO",
    month = "3",
    journal = {},
    year = "2026"
}

@incollection{Hill:2012,
  author    = {Hill, R. and Notte, J. A. and Scipioni, L.},
  title     = {},
  booktitle = {Advances in Imaging and Electron Physics},
  editor    = {Hawkes, P. W.},
  volume    = {170},
  pages     = {65--148},
  year      = {2012},
  publisher = {Elsevier},
  doi       = {10.1016/B978-0-12-394396-5.00002-6}
}

@article{Schaff:2015,
  author  = {Schaff, F. and Bech, M. and Zaslansky, P. and Jud, C. and Liebi, M. and Guizar-Sicairos, M. and Pfeiffer, F.},
  title   = {Six-dimensional real and reciprocal space small-angle X-ray scattering tomography},
  journal = {Nature},
  volume  = {527},
  pages   = {353--356},
  year    = {2015},
  doi     = {10.1038/nature16060}
}

@article{Baum:2019fqm,
    author = "Baum, Sebastian and Edwards, Thomas D. P. and Kavanagh, Bradley J. and Stengel, Patrick and Drukier, Andrzej K. and Freese, Katherine and G{\'o}rski, Maciej and Weniger, Christoph",
    title = "{Paleodetectors for Galactic supernova neutrinos}",
    eprint = "1906.05800",
    archivePrefix = "arXiv",
    primaryClass = "astro-ph.GA",
    reportNumber = "NORDITA-2019-060; LCTP-19-12",
    doi = "10.1103/PhysRevD.101.103017",
    journal = "Phys. Rev. D",
    volume = "101",
    number = "10",
    pages = "103017",
    year = "2020"
}

@incollection{GUO2012233,
  author    = {Guo, S.-L. and Chen, B.-L. and Durrani, S. A.},
  title     = {Solid-State Nuclear Track Detectors},
  booktitle = {Handbook of Radioactivity Analysis},
  edition   = {3},
  editor    = {L'Annunziata, M. F.},
  pages     = {233--298},
  year      = {2012},
  publisher = {Academic Press},
  address   = {Amsterdam},
  doi       = {10.1016/B978-0-12-384873-4.00004-9}
}

@article{Zhang:2026qnl,
    author = "Zhang, Xiuyuan and Thoyas, Andreas and Necib, Lina and Wetzel, Andrew and Arora, Arpit",
    title = "{Set the Night on FIRE: Building an Empirical Local Dark Matter Velocity Distribution}",
    eprint = "2603.25783",
    archivePrefix = "arXiv",
    primaryClass = "astro-ph.GA",
    month = "3",
    journal = {},
    year = "2026"
}

@article{Mao:2012hf,
    author = "Mao, Yao-Yuan and Strigari, Louis E. and Wechsler, Risa H. and Wu, Hao-Yi and Hahn, Oliver",
    title = "{Halo-to-Halo Similarity and Scatter in the Velocity Distribution of Dark Matter}",
    eprint = "1210.2721",
    archivePrefix = "arXiv",
    primaryClass = "astro-ph.CO",
    reportNumber = "SLAC-PUB-15326",
    doi = "10.1088/0004-637X/764/1/35",
    journal = "Astrophys. J.",
    volume = "764",
    pages = "35",
    year = "2013"
}

@article{Herzog-Arbeitman:2017fte,
    author = "Herzog-Arbeitman, Jonah and Lisanti, Mariangela and Madau, Piero and Necib, Lina",
    title = "{Empirical Determination of Dark Matter Velocities using Metal-Poor Stars}",
    eprint = "1704.04499",
    archivePrefix = "arXiv",
    primaryClass = "astro-ph.GA",
    doi = "10.1103/PhysRevLett.120.041102",
    journal = "Phys. Rev. Lett.",
    volume = "120",
    number = "4",
    pages = "041102",
    year = "2018"
}

@article{Necib:2018igl,
    author = "Necib, Lina and Lisanti, Mariangela and Garrison-Kimmel, Shea and Wetzel, Andrew and Sanderson, Robyn and Hopkins, Philip F. and Faucher-Gigu{\`e}re, Claude-Andr{\'e} and Kere{\v{s}}, Du{\v{s}}an",
    title = "{Under the Firelight: Stellar Tracers of the Local Dark Matter Velocity Distribution in the Milky Way}",
    eprint = "1810.12301",
    archivePrefix = "arXiv",
    primaryClass = "astro-ph.GA",
    doi = "10.3847/1538-4357/ab3afc",
    month = "10",
    journal = {},
    year = "2018"
}

@article{split_susy_conditions,
    author = "Giudice, G. F. and Romanino, A.",
    title = "{Split supersymmetry}",
    eprint = "hep-ph/0406088",
    archivePrefix = "arXiv",
    reportNumber = "CERN-PH-TH-2004-100",
    doi = "10.1016/j.nuclphysb.2004.08.001",
    journal = "Nucl. Phys. B",
    volume = "699",
    pages = "65--89",
    year = "2004",
    note = "[Erratum: Nucl.Phys.B 706, 487--487 (2005)]"
}

@misc{split_susy_little_hierarchy_2003,
    author = "Wells, James D.",
    title = "{Implications of supersymmetry breaking with a little hierarchy between gauginos and scalars}",
    booktitle = "{11th International Conference on Supersymmetry and the Unification of Fundamental Interactions}",
    eprint = "hep-ph/0306127",
    archivePrefix = "arXiv",
    reportNumber = "MCTP-03-30",
    month = "6",
    year = "2003"
}

@article{split_susy,
    author = "Arkani-Hamed, Nima and Dimopoulos, Savas",
    title = "{Supersymmetric unification without low energy supersymmetry and signatures for fine-tuning at the LHC}",
    eprint = "hep-th/0405159",
    archivePrefix = "arXiv",
    doi = "10.1088/1126-6708/2005/06/073",
    journal = "JHEP",
    volume = "06",
    pages = "073",
    year = "2005"
}

@article{LSP_1984,
title = {Supersymmetric relics from the big bang},
journal = {Nuclear Physics B},
volume = {238},
number = {2},
pages = {453-476},
year = {1984},
issn = {0550-3213},
doi = {https://doi.org/10.1016/0550-3213(84)90461-9},
url = {https://www.sciencedirect.com/science/article/pii/0550321384904619},
author = {John Ellis and J.S. Hagelin and D.V. Nanopoulos and K. Olive and M. Srednicki}
}

@article{Gonski:2025wzh,
    author = "Gonski, Julia L. and Graham, Peter W. and Rajendran, Surjeet and Ramani, Harikrishnan and Wong, Samuel S. Y.",
    title = "{Melting LHC detectors: a novel search for stopped long-lived particles}",
    eprint = "2512.12023",
    archivePrefix = "arXiv",
    primaryClass = "hep-ph",
    reportNumber = "FERMILAB-PUB-25-1005-SQMS-V",
    month = "12",
    year = "2025",
    journal = {}
}

@article{laurionite_conductivity,
author = {Natarajan, Mahadeva and Secco, Etalo A.},
title = {Studies on metal hydroxy compounds. XV. Electrical conductivity measurements on Cd(OH)Cl, Cu(OH)Cl, and PbOHCl},
journal = {Canadian Journal of Chemistry},
volume = {55},
number = {19},
pages = {3377-3379},
year = {1977},
doi = {10.1139/v77-474},
URL ={https://doi.org/10.1139/v77-474
},
eprint = {https://doi.org/10.1139/v77-474}
}

@article{LZ:2023lvz,
    author = "Aalbers, J. and others",
    collaboration = "LZ",
    title = "{First constraints on WIMP-nucleon effective field theory couplings in an extended energy region from LUX-ZEPLIN}",
    eprint = "2312.02030",
    archivePrefix = "arXiv",
    primaryClass = "hep-ex",
    doi = "10.1103/PhysRevD.109.092003",
    journal = "Phys. Rev. D",
    volume = "109",
    number = "9",
    pages = "092003",
    year = "2024"
}

@article{LZ:2022lsv,
    author = "Aalbers, J. and others",
    collaboration = "LZ",
    title = "{First Dark Matter Search Results from the LUX-ZEPLIN (LZ) Experiment}",
    eprint = "2207.03764",
    archivePrefix = "arXiv",
    primaryClass = "hep-ex",
    doi = "10.1103/PhysRevLett.131.041002",
    journal = "Phys. Rev. Lett.",
    volume = "131",
    number = "4",
    pages = "041002",
    year = "2023"
}

@article{ralph_mindatorg_2025,
	title = {Mindat.org: {The} open access mineralogy database to accelerate data-intensive geoscience research},
	volume = {110},
	issn = {0003-004X},
	url = {https://doi.org/10.2138/am-2024-9486},
	doi = {10.2138/am-2024-9486},
	number = {6},
	journal = {American Mineralogist},
	author = {Ralph, Jolyon and Von Bargen, David and Martynov, Pavel and Zhang, Jiyin and Que, Xiang and Prabhu, Anirudh and Morrison, Shaunna M. and Li, Wenjia and Chen, Weilin and Ma, Xiaogang},
	month = jun,
	year = {2025},
	note = {\_eprint: https://pubs.geoscienceworld.org/msa/ammin/article-pdf/110/6/833/7209752/ammin-2024-9486.1.pdf},
	pages = {833--844},
}

@misc{MindatLaurionite,
  author       = {{Mindat.org}},
  title        = {Laurionite: Mineral information, data and localities},
  year         = {2026},
  url          = {https://www.mindat.org/min-2343.html},
  note         = {Page updated March 2, 2026. Accessed April 1, 2026}
}

@article{regenspurg_formation_2016,
	title = {Formation and significance of laurionite in geothermal brine},
	volume = {75},
	issn = {1866-6299},
	url = {https://doi.org/10.1007/s12665-016-5668-4},
	doi = {10.1007/s12665-016-5668-4},
	number = {10},
	journal = {Environmental Earth Sciences},
	author = {Regenspurg, Simona and Driba, Dejene Legesse and Zorn, Carolin},
	month = may,
	year = {2016},
	pages = {865},
}

@misc{webmineral_nickel_strunz,
  title        = {Nickel-Strunz Classification},
  author = {{James A. Ferraiolo, Webmineral}},
  url          = {https://webmineral.com/strunz/strunz.php},
  urldate      = {2026-04-01},
  year={n.d.}
}

@book{strunz1941mineralogische,
  author    = {Strunz, Hugo},
  title     = {{Mineralogische Tabellen}},
  year      = {1941},
  address   = {Leipzig},
  publisher = {Akademische Verlagsgesellschaft Becker \& Erler}
}

@inbook{Sorrel1973Halides,
  author    = {Charles A. Sorrel},
  title     = {Halides},
  booktitle = {Rocks \& Minerals},
  pages     = {118--127},
  year      = {1973},
  publisher = {St Martin's Press},
  address   = {New York and Racine, WI},
  isbn      = {1-58238-124-0},
  note      = {Originally published as \emph{Minerals of the World}}
}

@article{Fleischer:1964,
  author  = {Fleischer, R. L. and Price, P. B. and Walker, R. M. and Hubbard, E. L.},
  title   = {Track Registration in Various Solid-State Nuclear Track Detectors},
  journal = {Physical Review},
  volume  = {133},
  pages   = {A1443--A1449},
  year    = {1964},
  doi     = {10.1103/PhysRev.133.A1443}
}

@article{Fleischer383,
  author  = {Fleischer, R. L. and Price, P. B. and Walker, R. M.},
  title   = {Tracks of Charged Particles in Solids},
  journal = {Science},
  volume  = {149},
  number  = {3682},
  pages   = {383--393},
  year    = {1965},
  doi     = {10.1126/science.149.3682.383}
}

@article{Fleischer:1965yv,
  author  = {Fleischer, R. L. and Price, P. B. and Walker, R. M.},
  title   = {Solid-State Track Detectors: Applications to Nuclear Science and Geophysics},
  journal = {Annual Review of Nuclear Science},
  volume  = {15},
  pages   = {1--28},
  year    = {1965},
  doi     = {10.1146/annurev.ns.15.120165.000245}
}

@techreport{Stewart1963MarineEvaporites,
  author       = {Stewart, Frederick H.},
  title        = {Marine Evaporites},
  year         = {1963},
  institution  = {U.S. Geological Survey},
  series       = {Professional Paper},
  number       = {440-Y},
  pages        = {Y1--Y52},
  doi          = {10.3133/pp440Y}
}

@misc{wyllie1970ultramafic,
  author    = {Peter J. Wyllie},
  title     = {Ultramafic rocks and the upper mantle},
  booktitle = {Fiftieth Anniversary Symposia: Mineralogy and Petrology of the Upper Mantle -- Sulfides -- Mineralogy and Geochemistry of Non-Marine Evaporites},
  editor    = {Benjamin A. Morgan},
  pages     = {3--32},
  year      = {1970},
  publisher = {Mineralogical Society of America},
  series    = {Special Paper},
  volume    = {3},
  url       = {https://msaweb.org/wp-content/uploads/2022/07/MSA_SP3_003-032.pdf},
  urldate   = {2026-04-02}
}

@incollection{KharakaHanor2003,
  author    = {Kharaka, Y. K. and Hanor, J. S.},
  title     = {Deep Fluids in the Continents: I. Sedimentary Basins},
  booktitle = {Treatise on Geochemistry},
  chapter   = {5.16},
  pages     = {1--48},
  year      = {2003},
  doi       = {10.1016/B0-08-043751-6/05085-4}
}

@article{Regenspurg2010,
  author  = {Regenspurg, Simona and Wiersberg, Thomas and Brandt, Wulf and Huenges, Ernst and Saadat, Ali and Schmidt, Katja and Zimmermann, G{\"u}nter},
  title   = {Geochemical properties of saline geothermal fluids from the in-situ geothermal laboratory Gro{\ss} Sch{\"o}nebeck (Germany)},
  journal = {Chemie der Erde - Geochemistry},
  volume  = {70},
  number  = {Suppl. 3},
  pages   = {3--12},
  year    = {2010},
  doi     = {10.1016/j.chemer.2010.05.002}
}

@article{Regenspurg2014,
  author  = {Regenspurg, Simona and Dilling, J{\"o}rg and Mielcarek, J{\"u}rgen and Korte, Frank and Schkade, Uwe-Karsten},
  title   = {Naturally occurring radionuclides and their geochemical interactions at a geothermal site in the North German Basin},
  journal = {Environmental Earth Sciences},
  volume  = {72},
  number  = {10},
  pages   = {4131--4140},
  year    = {2014},
  doi     = {10.1007/s12665-014-3306-6}
}

@article{de_marcillac_experimental_2003,
	title = {Experimental detection of {$\alpha$}-particles from the radioactive decay of natural bismuth},
	volume = {422},
	issn = {1476-4687},
	url = {https://doi.org/10.1038/nature01541},
	doi = {10.1038/nature01541},
	number = {6934},
	journal = {Nature},
	author = {de Marcillac, Pierre and Coron, Noël and Dambier, Gérard and Leblanc, Jacques and Moalic, Jean-Pierre},
	month = apr,
	year = {2003},
	pages = {876--878},
}

@Article{Geothermal_fluid_NewZealand_2023,
AUTHOR = {Sajkowski, Lucjan and Turnbull, Rose and Rogers, Karyne},
TITLE = {A Review of Critical Element Concentrations in High Enthalpy Geothermal Fluids in New Zealand},
JOURNAL = {Resources},
VOLUME = {12},
YEAR = {2023},
NUMBER = {6},
ARTICLE-NUMBER = {68},
URL = {https://www.mdpi.com/2079-9276/12/6/68},
ISSN = {2079-9276},
DOI = {10.3390/resources12060068}
}

@misc{anthony_handbook_mineralogy,
  author       = {John W. Anthony and Richard A. Bideaux and Kenneth W. Bladh and Monte C. Nichols},
  title        = {Handbook of Mineralogy},
  howpublished = {\url{http://www.handbookofmineralogy.org/}},
  publisher    = {Mineralogical Society of America},
  address      = {Chantilly, VA 20151-1110, USA}
}

@article{Kraemer_lowU_1981,
title = {234U and 238U concentration in brine from geopressured aquifers of the northern Gulf of Mexico basin},
journal = {Earth and Planetary Science Letters},
volume = {56},
pages = {210-216},
year = {1981},
issn = {0012-821X},
doi = {https://doi.org/10.1016/0012-821X(81)90128-X},
url = {https://www.sciencedirect.com/science/article/pii/0012821X8190128X},
author = {Thomas F. Kraemer}
}

@article{lowU_calculation_Langmuir_1978,
title = {Uranium solution-mineral equilibria at low temperatures with applications to sedimentary ore deposits},
journal = {Geochimica et Cosmochimica Acta},
volume = {42},
number = {6, Part A},
pages = {547-569},
year = {1978},
issn = {0016-7037},
doi = {https://doi.org/10.1016/0016-7037(78)90001-7},
url = {https://www.sciencedirect.com/science/article/pii/0016703778900017},
author = {Donald Langmuir}
}

@book{geochemistry_Krauskopf_1967,
  author    = {K. B. Krauskopf},
  title     = {Introduction to Geochemistry},
  publisher = {McGraw-Hill},
  address   = {New York, N.Y.},
  year      = {1967}
}

@misc{uranium_solubility_Goodwin_1980,
  author      = {B. W. Goodwin},
  title       = {Maximum total uranium solubility under conditions expected in a nuclear waste vault},
  institution = {Environmental and Safety Assessment Branch, Whiteshell Nuclear Research Establishment},
  number      = {TR-29},
  address     = {Pinawa, Manitoba},
  year        = {1980}
}

@article{McKibben1987SaltonSea,
  author  = {McKibben, M. A. and Williams, A. E. and Elders, W. A. and Eldridge, C. S.},
  title   = {Saline brines and metallogenesis in a modern sediment-filled rift: the Salton Sea geothermal system, California, U.S.A.},
  journal = {Applied Geochemistry},
  year    = {1987},
  volume  = {2},
  number  = {5--6},
  pages   = {563--578},
  doi     = {10.1016/0883-2927(87)90009-6}
}

@inproceedings{Scheiber2013ScalingInhibitorSoultz,
  author    = {Scheiber, J. and Seibt, A. and Birner, J. and Genter, A. and Moeckes, W.},
  title     = {Application of a Scaling Inhibitor System at the Geothermal Power Plant in Soultz-sous-For{\^e}ts: Laboratory and On-Site Studies},
  booktitle = {European Geothermal Congress EGC},
  year      = {2013},
  url       = {https://www.semanticscholar.org/paper/Application-of-a-Scaling-Inhibitor-System-at-the-in-Scheiber-Seibt/bbfb504bdbacc840c0597404e21b15b9e36a23cb}
}

@article{galena_and_laurionite_precipitate_2021,
title = {Characterization of scaling material obtained from the geothermal power plant of the Balmatt site, Mol},
journal = {Geothermics},
volume = {94},
pages = {102090},
year = {2021},
issn = {0375-6505},
doi = {https://doi.org/10.1016/j.geothermics.2021.102090},
url = {https://www.sciencedirect.com/science/article/pii/S037565052100050X},
author = {Jente Pauwels and Sonia Salah and Mirela Vasile and Ben Laenen and Valérie Cappuyns}
}

@article{Edwards_Gillard_Williams_Pollard_1992,
title={Studies of secondary mineral formation in the PbO-H2O-HC1 system},
volume={56},
DOI={10.1180/minmag.1992.056.382.07}, number={382},
journal={Mineralogical Magazine}, author={Edwards, R. and Gillard, R. D. and Williams, P. A. and Pollard, A. M.},
year={1992},
pages={53–65}}

@article{kolbel_waterrock_2020,
	title = {Water–rock interactions in the {Bruchsal} geothermal system by {U}–{Th} series radionuclides},
	volume = {8},
	issn = {2195-9706},
	url = {https://doi.org/10.1186/s40517-020-00179-4},
	doi = {10.1186/s40517-020-00179-4},
	number = {1},
	journal = {Geothermal Energy},
	author = {Kölbel, Lena and Kölbel, Thomas and Maier, Ulrich and Sauter, Martin and Schäfer, Thorsten and Wiegand, Bettina},
	month = sep,
	year = {2020},
	pages = {24},
}

@article{brine_precipitate_review_2025,
	title = {A review of recent advances in mineral scaling  in geothermal energy systems: mechanisms, mitigation, and case studies},
	volume = {84},
	issn = {1866-6299},
	url = {https://doi.org/10.1007/s12665-025-12416-9},
	doi = {10.1007/s12665-025-12416-9},
	number = {14},
	journal = {Environmental Earth Sciences},
	author = {Hassani, Kamran and Zheng, Wenbo},
	month = jul,
	year = {2025},
	pages = {418},
}

@article{laurionite_clogging,
title = {Hydraulic history and current state of the deep geothermal reservoir Groß Schönebeck},
journal = {Geothermics},
volume = {63},
pages = {27-43},
year = {2016},
note = {Enhanced Geothermal Systems: State of the Art},
issn = {0375-6505},
doi = {https://doi.org/10.1016/j.geothermics.2015.07.008},
url = {https://www.sciencedirect.com/science/article/pii/S0375650515000917},
author = {Guido Blöcher and Thomas Reinsch and Jan Henninges and Harald Milsch and Simona Regenspurg and Juliane Kummerow and Henning Francke and Stefan Kranz and Ali Saadat and Günter Zimmermann and Ernst Huenges}
}

@article{Kraemer_lowU_1986,
title = {Uranium geochemistry in geopressured-geothermal aquifers of the U.S. Gulf Coast},
journal = {Geochimica et Cosmochimica Acta},
volume = {50},
number = {6},
pages = {1233-1238},
year = {1986},
issn = {0016-7037},
doi = {https://doi.org/10.1016/0016-7037(86)90406-0},
url = {https://www.sciencedirect.com/science/article/pii/0016703786904060},
author = {Thomas F Kraemer and Yousif K Kharaka}
}

@techreport{geopressure_review_1992,
  author      = {Jane Negus-de Wys},
  title       = {The Geopressured Habitat: A Literature Review},
  institution = {Idaho National Engineering Laboratory},
  organization = {EG\&G Idaho, Inc.},
  address     = {Idaho Falls, ID},
  number      = {EGG-EP-9982},
  month       = sep,
  year        = {1992},
  note        = {Prepared for the U.S. Department of Energy},
  url         = {https://www.osti.gov/servlets/purl/896516}
}

@ARTICLE{matched_filter_1960,
  author={Turin, G.},
  journal={IRE Transactions on Information Theory}, 
  title={An introduction to matched filters}, 
  year={1960},
  volume={6},
  number={3},
  pages={311-329},
  doi={10.1109/TIT.1960.1057571}}

@article{DarkSide-50:2022qzh,
    author = "Agnes, P. and others",
    collaboration = "DarkSide-50",
    title = "{Search for low-mass dark matter WIMPs with 12~ton-day exposure of DarkSide-50}",
    eprint = "2207.11966",
    archivePrefix = "arXiv",
    primaryClass = "hep-ex",
    reportNumber = "FERMILAB-PUB-22-589-ND-PPD-SCD",
    doi = "10.1103/PhysRevD.107.063001",
    journal = "Phys. Rev. D",
    volume = "107",
    number = "6",
    pages = "063001",
    year = "2023"
}

@article{XENON:2023cxc,
    author = "Aprile, E. and others",
    collaboration = "XENON",
    title = "{First Dark Matter Search with Nuclear Recoils from the XENONnT Experiment}",
    eprint = "2303.14729",
    archivePrefix = "arXiv",
    primaryClass = "hep-ex",
    doi = "10.1103/PhysRevLett.131.041003",
    journal = "Phys. Rev. Lett.",
    volume = "131",
    number = "4",
    pages = "041003",
    year = "2023"
}

@article{PandaX:2024qfu,
    author = "Bo, Zihao and others",
    collaboration = "PandaX",
    title = "{Dark Matter Search Results from 1.54{\,}{\,}Tonne{\textperiodcentered}Year Exposure of PandaX-4T}",
    eprint = "2408.00664",
    archivePrefix = "arXiv",
    primaryClass = "hep-ex",
    doi = "10.1103/PhysRevLett.134.011805",
    journal = "Phys. Rev. Lett.",
    volume = "134",
    number = "1",
    pages = "011805",
    year = "2025"
}

@article{luminous_complex_WIMP,
    author = "Graham, Peter W. and Luengas Medina, Juan Pablo and Ramani, Harikrishnan and Tinoco Villalba, Daniel Antonio and Wong, Samuel S.~Y.",
    title = "(forthcoming)",
    journal="",
    year=2026
}

@article{Calabrese-Day:2026soq,
    author = "Calabrese-Day, Andrew and LaVoie-Ingram, Emilie and Ream, Kathryn and Ross, Hannah and Spitz, Joshua and Stengel, Patrick and Sun, Kai and Takla, Alexander",
    title = "{Toward Neutrino and Dark Matter Detection with Ancient Minerals: TEM Study of Heavy-Ion Tracks in Olivine}",
    eprint = "2604.09732",
    archivePrefix = "arXiv",
    primaryClass = "physics.ins-det",
    month = "4",
    year = "2026",
    journal = {},
}

@article{paleo_white_paper_2023,
    author = "Baum, Sebastian and others",
    title = "{Mineral detection of neutrinos and dark matter. A whitepaper}",
    eprint = "2301.07118",
    archivePrefix = "arXiv",
    primaryClass = "astro-ph.IM",
    reportNumber = "FERMILAB-PUB-23-501-SQMS-V",
    doi = "10.1016/j.dark.2023.101245",
    journal = "Phys. Dark Univ.",
    volume = "41",
    pages = "101245",
    year = "2023"
}

@article{Hedges:2026pgf,
    author = "Hedges, Samuel and Huber, Patrick",
    title = "{Calorimetric approach to paleo-detection of dark matter}",
    journal="",
    eprint = "2605.13659",
    archivePrefix = "arXiv",
    primaryClass = "hep-ph",
    month = "5",
    year = "2026"
}

\end{document}